\documentclass[preprint,authoryear,11pt]{elsarticle}  
\usepackage{fullpage}
\usepackage{graphicx}
\usepackage{amssymb}
\usepackage{amsmath}
\usepackage{multirow}
\usepackage{lineno}
\usepackage{subfig}

\usepackage{color}
\usepackage{soul}
\usepackage{array}
\usepackage{hyperref}
\usepackage[colorinlistoftodos]{todonotes}
\usepackage[]{algorithm2e}
\usepackage{varwidth}

\usepackage{changes}
\journal{}

\begin{document}

\begin{frontmatter}
\title{Particle dynamics in self-generated bedforms over a range of hydraulic
and sediment transport conditions using LES--DEM}

\author[1]{Rui Sun} \ead{sunrui@vt.edu}
\author[1]{Heng Xiao\corref{corhl}} \ead{hengxiao@vt.edu}
\address[1]{Department of Aerospace and Ocean Engineering, Virginia Tech, Blacksburg, Virginia, USA}

\author[2]{Kyle Strom} \ead{strom@vt.edu}
\address[2]{Department of Civil and Environmental Engineering, Virginia Tech, Blacksburg, Virginia, USA.}

\cortext[corhl]{Corresponding author. Tel: +1 540 231 0926}




\begin{abstract}
Direct measurement of vertical and longitudinal sediment fluxes on migrating sandy bedforms are
extremely difficult to perform in both the field and laboratory. In this study we use the LES--DEM
(large eddy simulation--discrete element method) solver \emph{SediFoam} to examine the individual
particle motions and resulting fluxes in a domain of self-generated bedforms. In \emph{SediFoam},
the motions of, and collisions among, the sediment grains as well as their interactions with
surrounding turbulent flows are captured. The numerical simulations are performed over a range of
transport settings, spanning bedform inception through washout conditions, to examine the individual
particle dynamics. The space-time evolution of sand bed surfaces is demonstrated. The self-generated
bedforms are stable at relatively low Reynolds numbers, but then become increasingly unstable at
higher Reynolds numbers; eventually washing out as the number of bypass grains and particles in
suspension increase.  Data from the simulation are used to examine the vertical entrainment rate of
particles and the fractionation of total sediment load into bed and suspended fractions as a
function of transport conditions. The study also compares the sediment transport rate obtained using
the bedform geometry and celerity to the true transport rate at different transport stages.
\end{abstract}

\end{frontmatter}

\section{Introduction}

The development of asymmetric bedforms at the interface of a flowing river and its bed is a
distinguishing feature of sand-bed rivers. These features arise naturally as dynamically stable
entities birthed from instabilities in the coupled fluid and sediment system, and their presence
strongly influences river hydraulic and sediment transport characteristics. Due to
the importance of dunes in river mechanics, and their compelling physical and mathematical
ascetics, their study has garnered substantial attention from researchers over many years
\citep[e.g.,][]{Gilbert1914, Anderson1953, Kennedy1963, MohrigSmith1996, McElroyMohrig2009,
CharruEtal2013}. Nevertheless, there is still much about dunes we are currently learning
\citep[e.g.,][]{NaqshbandEtal2014, naqshband15mr, emadzadeh16ms, VendittiEtal2016}, and a unified
quantitative approach to account for sediment mass flux in the longitudinal and vertical
directions in the presence of migrating dunes is lacking \citep{Garcia2008sedman}.

\begin{figure}[h] 
\centering
\includegraphics[width=0.45\textwidth,angle=0]{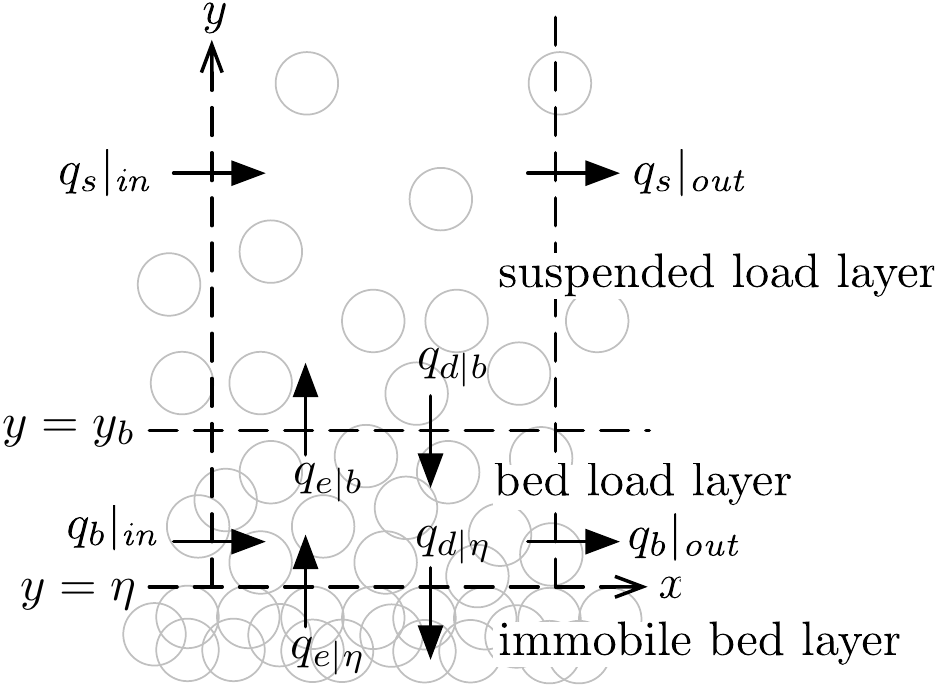}
\caption{Sketch of the three-layer conceptualization of a mobile sediment bed.}
\label{fig:layers}
\end{figure}

Classic modeling of sediment transport in sand bed rivers conceptually discretizes
  active river beds into three vertical layers based on particle mobility and mode of transport
  (Fig.~\ref{fig:layers}).  These three layers consist of a stable bed layer (with particle
  velocities, $u_{p}=0$, and solids volume fraction, or volume concentration, $\varepsilon_s$,
  greater than 0.5, and a vertical coordinate $y<\eta$), a mobile, but dense, bed load layer with
  variable thickness, $\delta_{s}$, ($\eta\leq y \leq y_{b}$), and a more disperse suspended phase
  ($y>y_{b}$); based on this particular definition, the elevation of the bed surface, $\eta$, is
  taken to coincide with the top of the immobile layer. The unit width volume flow rate of sediment
  traveling between $y=\eta$ and $y=y_{b}$ is the bed load transport rate, $q_{b}$ $[L^{2}t^{-1}]$,
  and the volume flow rate of sediment moving above the elevation of $y=y_{b}$ is the suspended
  load, $q_{s}$ $[L^{2}t^{-1}]$;  the summation of the bed and suspended loads is the total load,
  $q_{t}$. Exchange between the bed and bed load layer is accounted for with the entrainment and
  deposition rates evaluated at $y=\eta$, $q_{e|\eta}$ and $q_{d|\eta}$. Similarly exchange between
  the bed load and suspended layers is accounted for using erosion and deposition functions
  evaluated at $y=y_{b}$, $q_{e|b}$ and $q_{d|b}$ (Fig.~\ref{fig:layers}). 

In typical reach-scale sediment transport calculations and morphodynamic simulations, the individual
sediment particles are not directly resolved; nor are, in most cases, ripple or dune-scale changes
in the bed elevation~\citep{giri06nc,khosronejad14ns}. Instead, sediment is modeled as a continuous
substance, and moved throughout the domain using the vertical exchange terms, the computed fluid
velocities, and the longitudinal bed or total loads. Common assumptions are that material in
suspension travels with a speed equal to the fluid velocity; the bed load is in equilibrium with
local conditions; and that $q_{d|b}$ and $q_{e|b}$ can be modeled as:
$q_{d|b}=w_{s}\varepsilon_{s|b}$ and $q_{e|b}=w_{s}E_{s}=w_{s}\varepsilon_{s|be}$, where
$w_{s}$ is the terminal settling velocity of the sediment, $\varepsilon_{s|b}$ is sediment
concentration near the interface of the suspended and bed load layer, and $E_{s}$ (or
$\varepsilon_{s|be}$) is the entrainment function, which is taken to be the near bed concentration
under equilibrium conditions when erosion and deposition are locally balanced. Therefore, to model
the movement of sediment and the resulting bed deformation, the closure equations for the sediment
phase that are needed are the bed load flow rate, $q_{b}$, and the entrainment function, $E_{s}$.

Many empirical functions for these two quantities have been proposed over the years
\citep{Garcia2008sedman}. However, obtaining data to evaluate $q_{b}$, $q_{s}$, and $E_{s}$
independently and in detail is very difficult. Rather than measuring individual particle movements,
sediment loads are measured in mass over some time interval and often only at one particular
location along a stretch of river or laboratory flume. Furthermore, disentangling $q_{t}$, $q_{s}$,
and $q_{b}$ from experimental data, which typically consist of measurements of $q_{t}$ and
$\varepsilon_s$, can be difficult, and no direct measurements of $E_{s}$, that we are aware of, have
been made in cases of actively migrating bedforms. Rather, reach averaged values of $E_{s}$ are
extracted from the vertical concentration profiles under equilibrium conditions \citep{Guy1966,
GarciaParker1991}.
  
When thinking about sand beds with migrating dunes, it is clear that spatial heterogeneities in the
bed elevations will drive heterogeneities in the longitudinal and vertical turbulence properties of
the flow and, hence, particle mobility \citep{SmithMclean1977, vanRijn1984, Best2005a}. For example,
the recent study of \cite{emadzadeh16ms} has highlighted that sediment entrainment rates vary at
different spatial locations over fixed idealized dunes. While most sediment transport calculations
do not resolve bedform scale features, it is possible that further insights into the mechanics
associated with actively migrating ripples and dunes in turbulent flows may come from a better
understanding of particle motion over the bedform. In addition, the translation of these mechanics
to sediment transport models such as those for $q_{b}$ and $E_{s}$, could potentially be enhanced by
examining the details of individual particle motions.

For example, advances have recently been made in understanding bed load transport in gravel beds
using particle statistics obtained from high-speed cameras \citep{Ancey2010, Lajeunesseetal2010b,
Roseberryetal2012, CampagnolEtal2015, FathelEtal2015}. Concurrent with this new data has been a
drive to better describe the motion of the discrete sediment phase in terms of continuous river
properties that can be used in larger-scale morphodynamic modeling \citep{Parkeretal2000, Ancey2010,
Lajeunesseetal2010b, FurbishEtal2012a, FurbishEtal2016}. The focus of these recent studies has been
on the more sporadic motion of gravel particles traveling over a plane flume bed under equilibrium
transport conditions with steady, uniform time-averaged hydraulics. Such conditions are different
from those producing actively migrating sandy bedforms, but it is likely that some of the basic
methodology and discussion generated by this body of work could also help to improve understanding
of transport in sand bed rivers. 

Examining particle motion in the presence of translating sandy bedforms presents at least two
complications. The first is that steady hydraulics can only be obtained within a frame of reference
that translates with the bedform. Secondly, even within a moving frame of reference, both vertical
and longitudinal variation in the fluid and transport characteristics exists -- even in the time
averaged quantities \citep{KostaschukEtal2009, NaqshbandEtal2014}. Furthermore, operations such as
tracking the jump length of sand grains in experimental ripples and dunes are difficult. This is due
to the combination of particle sizes smaller than gravels, travel distances that can be up to and
longer than one bedform wavelength, large numbers of nearly identical particles in motion, and
opaque water. Needless to say, such constraints exceed the limits that even the state of the art
imaging and image processing technology can currently achieve.

A secondary approach to bolstering our understanding of particle motion in beds with actively
migrating bedforms, is to make use of interface-resolved direct numerical simulations, which
resolve both the turbulent flow, the flow field surrounding each particle, and the interparticle
collisions~\citep{AussillousEtal2013, FukuokaEtal2014, KidanemariamUhlmann2014}. Such methods have
proved especially fruitful in painting a better picture of the interplay between turbulence and
sediment transport over plane beds \citep[e.g.,][]{KempeEtal2014, VowinckelEtal2014}. A drawback to
such methods, however, is that they are highly expensive computationally. For example, the
computational cost for sediment transport simulation of 27,000 grains on a flat bed is 30 million
CPU hours~\citep{VowinckelEtal2014}. Because of the expense, most simulations have focused on active
sediment transport in the absence of self-generated bedforms; an exception to this statement is the
study of \cite{KidanemariamUhlmann2014}.

In the past few years researchers started to use modern, general-purpose particle-resolving solvers
based on LES--DEM to study sediment transport. The LES--DEM has been used extensively in the past
two decades in the chemical and pharmaceutical industry on a wide range of applications such as
fluidized beds, cyclone separator, and pneumatic conveying~\citep{han03DEM,ebrahimi14cfd}.  In
LES--DEM, Large eddy simulation is used to model the fluid flow, which is coupled with the discrete
element method for resolving particle motions. In his pioneering work, \cite{Schmeeckle2014} used an
open-source LES--DEM solver~\citep{goniva09tf,kloss12ma} to study suspended sediment transport over
featureless beds. The LES--DEM approach is also proved to be capable of simulating the
self-generated ripples/dunes in a pipe or channel \citep{arolla15tm,SunXiao2016b}. 
It should be noted that the hydro- and morphodynamic models for sediment transport
simulations also gained success in the prediction of dune generation and
migration~\citep{shimizu09ns,escauriaza11lm,nabi13ds2,nabi13ds3}. For example,
that~\cite{nabi13ds2,nabi13ds3} uses an LES model and a DEM approach based on complete particle
equations of motion.  In addition, the number of elements in the DEM model is scaled-each
particles accounts for more mass flux than a single sediment particle, such that the physics are
retained but the computational load is reduced. However, the computation of the sediment erosion
in Nabi's work is still based on the empirical correlations to describe sediment erosion and
deposition fluxes. Compared with the hydro- and morphodynamic models, the LES--DEM approach resolves
the motion of individual particles, which is an important improvement.

It has recently been shown that \emph{SediFoam}, a hybrid LES--DEM solver for particle-laden flows
\citep{SunXiao2016a}, is capable of producing self-organized bedforms with characteristic height and
migration celerities similar to those observed in laboratory flumes \citep{SunXiao2016b}.
In this study, our first aim is to use \emph{SediFoam} to examine the individual
particle dynamics of sand grains in a field of self-organize asymmetric ripples. This is done over a
range of transport settings, spanning from bedform inception to washout conditions. The particle
velocities are then temporally and spatially averaged in a moving frame of reference to extract key
sediment transport quantities. We also compared the averaged values to empirical formulas used in
one-dimensional river morphodynamic modeling, with the primary intent of showing that DEM approaches
can be used to extract the types of data needed to improve understanding and modeling in cases of
actively migrating sandy bedforms. Examining these properties in a moving frame of reference also
allows us to examine the link between translation and deformation of ripples/dunes in light of
spatially varying bed and suspended loads.

\section{Methodology}
\label{sec:lesdem}

\emph{SediFoam} is an open-source, highly parallel, robust LES--DEM solver with emphasis on sediment
transport applications. It is capable of simulating the interaction between turbulent structures and
sediment particles. The modeling of translational and rotational motion of each sediment particle is
based on Newton's second law as the following equations~\citep{cundall79}: 
\begin{subequations}
 \label{eq:newton}
 \begin{align}
  m \frac{d\mathbf{u}_p}{dt} &
  = \mathbf{f}^{col} + \mathbf{f}^{fp} + m \mathbf{g} \label{eq:newton-v}, \\
  I \frac{d\mathbf{\Psi}_p}{dt} &
  = \mathbf{T}^{col} + \mathbf{T}^{fp} \label{eq:newton-w},
 \end{align}
\end{subequations}
where $m$ is the mass of particle; \( \mathbf{u}_p \) is particle velocity; $t$ is time;
\(\mathbf{f}^{col} \), \(\mathbf{f}^{fp}\) are particle--particle collision force and
fluid--particle interaction force of Lagrangian particles, respectively;
\(\mathbf{g}\) denotes gravity. Similarly, \(I\) and \(\mathbf{\Psi}_p\) are angular moment of
inertia and angular velocity of the particle; \(\mathbf{T}^{col}\) and \(\mathbf{T}^{fp}\) are the
torques due to inter-particle contact and fluid--particle interaction, respectively. Note that the
vectors and tensors are denoted using boldface letters (e.g., \(\mathbf{u}\)). To compute the
collision forces and torques, the particles are modeled as soft spheres with inter-particle contact
represented by an elastic spring and a viscous dashpot. Further details can be found
in~\cite{tsuji93}.

Locally-averaged incompressible Navier--Stokes equations are used to describe the fluid flow by
applying local averaging the variables such as fluid velocity and pressure.  Assuming constant fluid
density \(\rho_f\), the governing equations for the fluid are~\citep{anderson67,kafui02}:
\begin{subequations}
 \label{eq:NS}
 \begin{align}
  \nabla \cdot \left(\varepsilon_s \mathbf{U}_s + {\varepsilon_f \mathbf{U}_f}\right) &
  = 0 , \label{eq:NS-cont} \\
  \frac{\partial \left(\varepsilon_f \mathbf{U}_f \right)}{\partial t} + \nabla \cdot \left(\varepsilon_f \mathbf{U}_f \mathbf{U}_f\right) &
  = \frac{1}{\rho_f} \left( - \nabla p + \varepsilon_f \nabla \cdot \mathbf{\tau} + \varepsilon_f \rho_f \mathbf{g} + \mathbf{F}^{fp}\right), \label{eq:NS-mom}
 \end{align}
\end{subequations}
where \(\varepsilon_s\) is the solid volume fraction; \( \varepsilon_f = 1 - \varepsilon_s \) is the
fluid volume fraction; \( \mathbf{U}_f \) is the fluid velocity. The terms on the right hand side of
the momentum equation are: pressure gradient \(\nabla p\), divergence of the stress tensor \(
\mathbf{\tau} \) (including viscous and Reynolds stresses), gravity, and fluid--particle interaction
forces per unit mass, respectively.
Note that the velocities and the forces associated with individual particles are denoted as
$\mathbf{u}$ and $\mathbf{f}$, respectively; and the velocities and forces in the continuum scale
are denoted as $\mathbf{U}$ and $\mathbf{F}$, respectively.
In the present study, large-eddy simulation is
applied to resolve the flow turbulent motions that are larger than the spatial filter which is the
same as the grid size. The stress tensor is composed of both viscous and Reynolds stresses: \(
\mathbf{\tau} = \mu\nabla \mathbf{U}_f + \rho_f \mathbf{\mathcal{R}}\), where $\mu$ is the molecular
viscosity of fluid and $\mathbf{\mathcal{R}}$ is the Reynolds stress. The expression of the Reynolds
stress is:
\begin{equation}
  \mathbf{\mathcal{R}} = \frac{2}{\rho_f}\mu_t \mathbf{S} - \frac{2}{3}k\mathbf{I},
  \label{eq:reynolds-stress}
\end{equation}
where $\mu_t$ is the eddy viscosity, $\mathbf{S} = (\nabla \mathbf{U}_f + (\nabla
\mathbf{U}_f)^T)/2$ is the rate-of-strain tensor, $k$ is the turbulent kinetic energy, and $I$ is
the identity matrix.  We applied the one-equation eddy viscosity model proposed
by~\cite{yoshizawa85sd} as the sub-grid scale (SGS) model. The one-equation eddy model solves a
transport equation for the sub-grid scale kinetic energy on which the eddy viscosity depends. It is
observed by~\cite{yoshizawa85sd} that the standard Smagorinsky model would be recovered from the
one-equation eddy model if production equals dissipation.  However, the one-equation eddy model can
provide better accuracy under non-equilibrium circumstances in channel
flows~\citep{fureby97les,de07tp}. It is noted that in the stress tensor \( \mathbf{\tau} \) term,
the fluctuations of the fluid flow at the boundary of the particle are not resolved. The Eulerian
fields $\varepsilon_s$, $\mathbf{U}_s$, and $\mathbf{F}^{fp}$ in Eq.~(\ref{eq:NS}) are obtained
based on the volume, velocity, and fluid-particle interaction force from each Lagrangian particle
contained in a given fluid grid cell.

The fluid-particle interaction force \(\mathbf{f}^{fp}\) in Eq.~(\ref{eq:newton-v}) consists of
buoyancy \( \mathbf{f}^{buoy} \), drag \( \mathbf{f}^{drag} \), lift force \(\mathbf{f}^{lift}\),
and pressure gradient force \( \mathbf{f}^{pg} \). The drag force model proposed
by~\cite{mfix93} is applied to the present simulations. The drag on an individual component sphere
$i$ is formulated as:
\begin{equation}
  \mathbf{f}^{drag}_i = \frac{V_{p,i}}{\varepsilon_{f, i} \varepsilon_{s, i}} \beta_i \left(
  \mathbf{U}_{f, i} - \mathbf{u}_{p,i}
 \right),
  \label{eqn:particleDrag}
\end{equation}
where \( V_{p, i} \) and \( \mathbf{u}_{p, i} \) are the volume and the velocity of particle $i$,
respectively; \( \mathbf{U}_{f, i} \), \( \varepsilon_{f, i} \), and \( \varepsilon_{f, i} \) are
the fluid velocity, solid volume fraction, and fluid volume fraction interpolated to the center of
particle $i$, respectively; \( \beta_{i} \) is the drag correlation coefficient which accounts for
the presence of other particles. The $\beta_i$ value in the present study is based on~\cite{mfix93}:
\begin{equation}
  \beta_i = \frac{3}{4}\frac{C_{d,i}}{\Gamma_{r,i}^2} \frac{\rho_f |\mathbf{U}_{f,i}
  - \mathbf{u}_{p,i}|}{d_{p,i}}\varepsilon_{f,i}\varepsilon_{s,i}
  \mathrm{, \quad with \quad} C_{d,i} = \left( 0.63+0.48\sqrt{\Gamma_{i}/\mathrm{Re_{pv,i}}} \right)^2,
  \label{eqn:beta-i}
\end{equation}
where $C_{d,i}$ is the drag coefficient of particle $i$; $d_{p,i}$ is the diameter of particle $i$;
the particle velocity Reynolds number $\mathrm{Re_{pv,i}}$ is defined as:
\begin{equation}
  \mathrm{Re_{pv,i}} = \rho_s d_{p,i}
  |\mathbf{U}_{f,i} - \mathbf{u}_{p,i}|/\mu;
  \label{eqn:p-re}
\end{equation}
the $\Gamma_{i}$ is the correlation term for the $i$-th particle:
\begin{equation}
  \Gamma_{i} = 0.5\left( A_{1,i} - 0.06\mathrm{Re_{pv,i}}+\sqrt{(0.06\mathrm{Re_{pv,i}})^2 +
  0.12\mathrm{Re_{pv,i}}(2A_{2,i} - A_{1,i}) + A_{1,i}^2} \right),
  \label{eqn:drag-vr}
\end{equation}
where $A_1$ and $A_2$ are asymptotic values of the relative velocity (i.e., the ratio between the
particle terminal velocity in the presence of other particles and the terminal velocity of a single
particle) at low and high Reynolds numbers, respectively~\citep{garside77vv}:
\begin{equation}
  A_{1,i} = \varepsilon_{f,i}^{4.14}, \quad
  A_{2,i} =
  \begin{cases}
  0.8\varepsilon_{f,i}^{1.28} & \quad \text{if } \varepsilon_{f,i} \le 0.85, \\
  \varepsilon_{f,i}^{2.65}    & \quad \text{if } \varepsilon_{f,i} > 0.85.\\
  \end{cases}
  \label{eqn:drag-A}
\end{equation}

In addition to drag, the lift force is considered in the simulations. The velocity gradient in the
flow field can lead to the lift forces on a spherical particle. This is because the velocity
gradient will result in a flow deflection at the particle surface so that the flow velocity, and
consequently the pressure, on one side can be different from the other side.  The pressure
difference causes the lift forces. The lift force on a spherical component particle is modeled
as~\citep{saffman65th,rijn84se1,zhu07dps}:
\begin{equation}
  \mathbf{f}_{i}^{lift} = C_{l} (\rho_f \mu)^{0.5} d_{p,i}^{2} \left( \mathbf{U}_{f,i} -
  \mathbf{u}_{p,i} \right) \mathbf{\times}
  \frac{\mathbf{\omega}_i}{|\mathbf{\omega}_i|^{0.5}},
  \label{eqn:particleLift}
  \end{equation}
where $C_{l} = 1.6$ is the lift coefficient, $\mathbf{\omega}_i = \nabla \times \mathbf{U}_{f,i}$ is
the flow vorticity interpolated to the center of particle $i$, and $\mathbf{\times}$ indicates the
cross product between vectors. The pressure gradient force \( \mathbf{f}^{pg} \) is
modeled as~\citep{chu09cfd}:
\begin{equation}
  \mathbf{f}_{i}^{pg} = V_{p, i} \nabla p_{i},
\label{eqn:particlePG}
\end{equation}
where $p_{i}$ is the pressure interpolated to the center of particle $i$. We consider the pressure
gradient force because temporal pressure fluctuations on the scale of the particles can
produce far larger instantaneous forces.

\section{Implementations and numerical methods}
\label{sec:num-method}
The hybrid LES--DEM solver \emph{SediFoam} is developed based on two state-of-the-art open-source
codes in their respective fields. LES is performed by using CFD (Computational Fluid Dynamics)
toolbox OpenFOAM (Open Field Operation and Manipulation)\citep{openfoam}, and DEM is performed by a
molecular dynamics simulator LAMMPS (Large-scale Atomic/Molecular Massively Parallel Simulator)
which is developed at the Sandia National Laboratories~\citep{lammps}. A parallel LAMMPS--OpenFOAM
interface is implemented for the communication of the two open-source solvers. The code is publicly
available at https://github.com/xiaoh/sediFoam under GPL license. Detailed
introduction of the implementations of this hybrid solver is discussed in~\cite{SunXiao2016a}.

The fluid equations in~(\ref{eq:NS}) are solved in OpenFOAM with the finite volume method
\citep{jasak96ea}. The coupled pressure and velocity fields are solved by using PISO (Pressure
Implicit Splitting Operation) algorithm~\citep{issa86so}. A second-order central scheme is used for
the spatial discretization of convection and diffusion terms. Time integrations are performed with a
second-order backward Euler scheme. An averaging algorithm based on diffusion is implemented to
obtain smooth $\varepsilon_s$, $\mathbf{U}_s$ and $\mathbf{F}^{fp}$ fields from discrete sediment
particles~\citep{sun15db1, sun15db2}. In the averaging procedure to get the averaged Lagrangian
particle quantities ($\varepsilon_s$, $\mathbf{U}_s$, and $\mathbf{F}^{fp}$ in Eq.~(\ref{eq:NS})),
the diffusion equations are solved on the same mesh for LES. A second-order central scheme is used
for the spatial discretization of the diffusion equation; a second-order implicit scheme is used for
the temporal integration. To resolve the collision between the sediment particles, the contact force
between sediment particles is computed with a linear spring-dashpot model~\citep{cundall79}. The
time step to resolve the particle collision is less than 1/50 the contact time to avoid particle
inter-penetration~\citep{sun07ht}. 

The LES--DEM solver \emph{SediFoam} has been verified and validated extensively with cases of
general particle-laden flows and specific applications in sediment transport. \cite{gupta15vv}
verified the implementation of the inter-phase momentum transfer term and validates the solver
against various fluidized bed applications; \cite{SunXiao2016a} validated \emph{SediFoam} by using a
test of 500 moving sphere particles and found the results agree well with the DNS
results~\citep{kempe12cm}; \cite{SunXiao2016b} simulated several different sediment transport
regimes, including ``flat bed in motion'', ``small dune'' and ``vortex dune'', and the results agree
with experiments~\citep{ouriemi09sd2} and interface resolving simulations
of~\cite{kidanemariam14laminar,KidanemariamUhlmann2014}; \cite{sun16micro} demonstrated the
capability of the solver in the simulation of sediment transport in sheet flows induced by
oscillatory boundary layer.  More recently, the feasibility of simulating irregular particles with
\emph{SediFoam} by using bonded spheres has also been demonstrated and the results were compared
with experimental measurements~\citep{sun16irregular}.  In this study, we performed an additional
validation of grain avalanching, the results of which are presented in the Appendix.  The repose
angle obtained in the avalanching test is consistent with those in the
literature~\citep{goniva12irf,zhao13coupled}.   Therefore, we can conclude that the capabilities of
\emph{SediFoam} in representing particle dynamics and fluid-particle interactions in various regimes
of sediment transport have been adequately demonstrated. The objective of the present study is to
use this well-validated high-fidelity simulation tool to probe the physics of particle dynamics in
self-generated bedforms.

\section{Conditions for runs}
The numerical tests of the self-generated bedforms are performed at different bulk Reynolds numbers
$\mathrm{Re_b}$ varying from 6,000 to 12,000. The bedforms grow in size from Case 1 to
Case 2, and then began to wash out in Cases 3 and 4. The bulk Reynolds number $\mathrm{Re_b}$ is
defined as $2 H_b U_b/\nu$, where $H_b = 23d_p$ is the equivalent boundary layer thickness as
measured from the initial bed surface to the top of the computational domain; $U_b$ is the mean bulk
flow velocity in the equivalent boundary layer; $\nu$ is the kinetic viscosity of the fluid. The
selection of the equivalent boundary layer thickness is according
to~\citep{KidanemariamUhlmann2014}. The thickness of the initial flat sediment bed layer is $L_y -
H_b = 10d_p$. The setup of the numerical tests is based on previous studies of bedform
migration~\citep{KidanemariamUhlmann2014,SunXiao2016b}. LES--DEM resolved four-way coupled
interactions among the fluid and particles, which include fluid-to-particle, particle-to-fluid, and
particle--particle interactions~\citep{kuipers97ac}. At each time step, the DEM gives the
information to update the fluid field, and the fluid-particle interaction force according to the
updated fluid flow data is incorporated into the DEM. The trajectories of the sediment particles are
captured by using DEM and thus we can investigate the characteristics of individual particles.

\begin{table}[htbp]
 \caption{Parameters of the numerical simulations.}
 \begin{center}
 \begin{tabular}{lcccc}
   \hline       & Case 1 & Case 2 & Case 3 & Case 4 \\
   Description  & small dune & large dune & washing out & washed out \\
   \hline
   Domain dimensions                            &\\
   \qquad length, height, thickness & \multicolumn{4}{c}{\multirow{2}{*}{$312\times33\times80$}} \\
   \qquad ($L_x/d_p$, $L_y/d_p$, $L_z/d_p$) & \multicolumn{4}{c}{} \\
   \qquad length, height, thickness & 
   \multicolumn{4}{c}{\multirow{2}{*}{$156\times16.5\times40$}} \\
   \qquad in mm ($L_x$, $L_y$, $L_z$) & \multicolumn{4}{c}{} \\
   Mesh resolutions                             &\\                        
   \qquad length, height, thickness ($N_x$, $N_y$, $N_z$)                         
   & \multicolumn{4}{ c }{$120\times140\times40$} \\
  Particle properties & \\
   \qquad total number                          & \multicolumn{4}{ c }{272,000}\\
   \qquad diameter $d_p$~[mm]                   & \multicolumn{4}{ c }{0.5}\\
   \qquad density $\rho_s$~[$\times 10^3~\mathrm{kg/m^3}$]  & \multicolumn{4}{ c }{2.5} \\
   \qquad terminal settling velocity $w_s$~[$\mathrm{m/s}$]  & \multicolumn{4}{ c }{0.06} \\
   \qquad particle stiffness coefficient~[N/m]  & \multicolumn{4}{ c }{200} \\
   \qquad normal restitution coefficient        & \multicolumn{4}{ c }{0.3} \\
   \qquad coefficient of friction               & \multicolumn{4}{ c }{0.4} \\
   Fluid properties and flow conditions & \\
   \qquad kinetic viscosity $\nu$~[$\times 10^{-6}~\mathrm{m^2/s}$] & \multicolumn{4}{ c }{1.5} \\
   \qquad density $\rho_f$~[$\times 10^3~\mathrm{kg/m^3}$]  & \multicolumn{4}{ c }{1.0} \\
   \qquad boundary layer thickness $2 H_b/d_p$  & \multicolumn{4}{c}{46} \\    
   \qquad bulk Reynolds number $\mathrm{Re_b} = 2 U_b H_b/\nu$       
                                                & 6,000  & 8,000  & 10,000 & 12,000 \\
   \qquad mean flow velocity $U_b$~[m/s]        & 0.39  & 0.52  & 0.65  & 0.78 \\
   \hline
  \end{tabular}
 \end{center}
 \label{tab:param-all}
\end{table}

\begin{figure}[htbp]
  \centering
  \subfloat[Layout]{
    \includegraphics[width=0.65\textwidth]{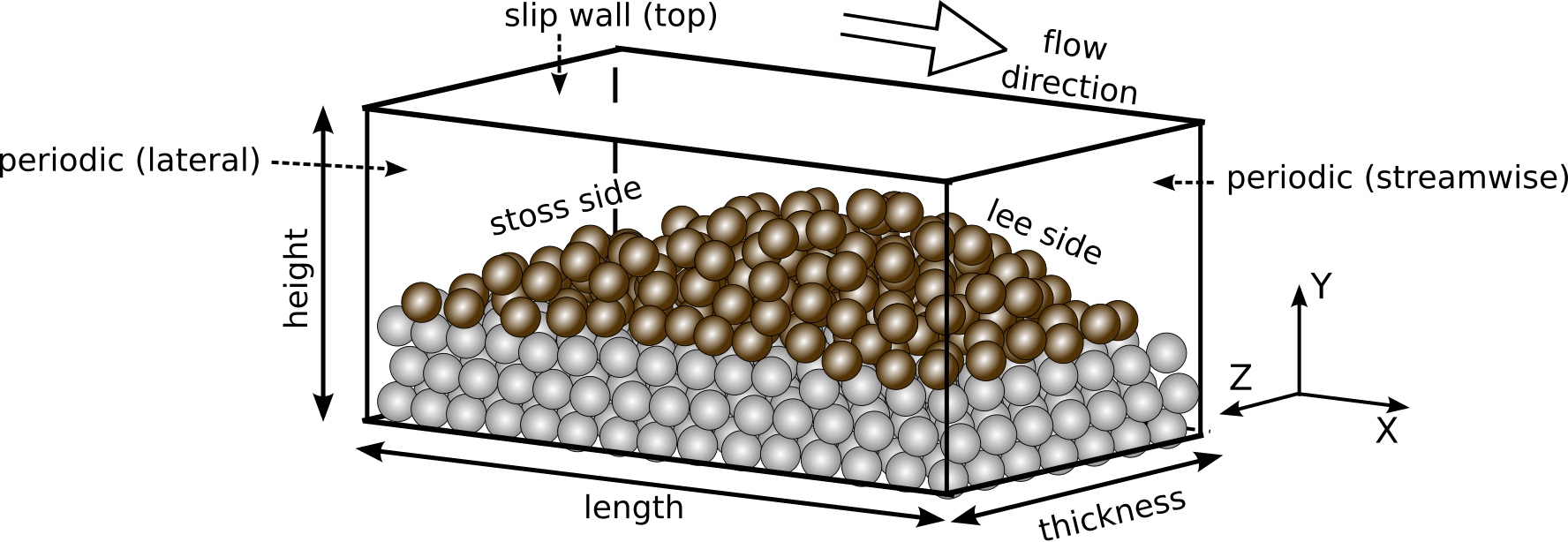}
  }
  \subfloat[Forces Computation]{
    \includegraphics[width=0.25\textwidth]{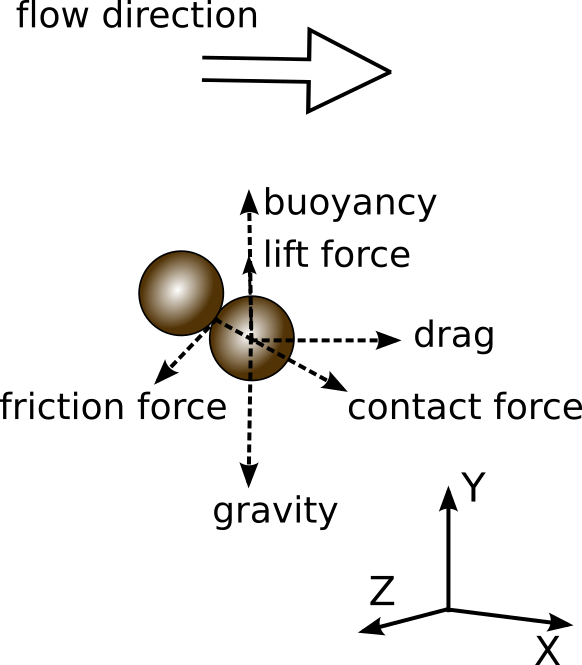}
  }
  \hspace{0.1in}
  \subfloat[Simulation Snapshot]{
    \includegraphics[width=0.65\textwidth]{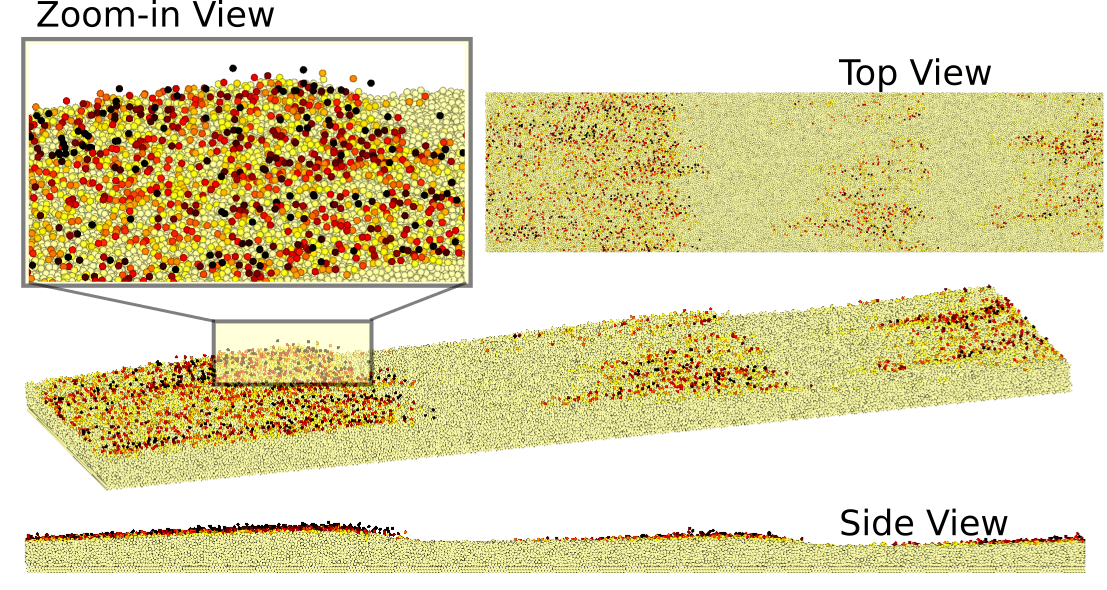}
  }
  \caption{Layout of the numerical simulations of self-generated bedforms (Panel A), the force
  computation on individual particle (Panel B), and a snapshot of the simulation (Case 1:
  $\mathrm{Re_b}$ = 6,000). In Panel A, the brown particles are moving particles; the gray particles
  are fixed to provide rough wall boundary conditions. Note that the particle and bedform sizes are
  not to scale. In Panel B, only the forces on one particle are plotted. In Panel C, the zoom-in
view, top view, and side view are also presented.}
  \label{fig:layout}
\end{figure}

The dimensions of the domain, the mesh resolutions, and the fluid and particle properties are
detailed in Table~\ref{tab:param-all}.  The geometry of the domain is shown in
Fig.~\ref{fig:layout}(a), and $x$-, $y$- and $z$- coordinates are aligned with the streamwise,
wall-normal, and lateral directions, respectively. It is important to note that the computational
domain sizes must be large than the size of the largest turbulent coherent structures and the
largest bedform patterns that may develop during the simulations. Clearly, in the simulations no
turbulent structures or bedform patterns large than the domain can develop, and an inadequate domain
size would artificially suppress the development of these structures.  On the other hand, as the
sediment particle diameter is given and the CFD mesh needs to be fine enough to resolve the smallest
energetic turbulent eddies, increasing the domain size would increase the total number of particles
and the number of grid cells in the simulations and thus would increase the costs.  Therefore, it is
important to use a domain size that is large enough but not much larger in consideration of reducing
computational cost.  A recent study of~\cite{kidanemariam17fsp} has found that the minimum length to
generate stable ripples is about 100$d_p$.  They further demonstrated that the dune features (i.e.,
height, wavelength, migration velocity) do not vary significantly when the computational domain
length is larger than the minimum length. In this present simulations we used domain size of
312$d_p$, which is larger than the minimum size required 100$d_p$. The physical
parameters of the simulation are also shown in Table~\ref{tab:param-all}.  The particle diameter is
0.5~mm and the channel height is 16.5~mm. The sizes of small dunes generated in Case 1 is about
80~mm after 24s. The bulk velocities in the simulations varies from 0.39 to 0.78~m/s. For Case 1,
the time step is $5\times10^{-4}$~s for the fluid flow and $1\times10^{-6}$~s for the particles.

The LES mesh in streamwise ($x$-) and lateral ($z$-) directions is uniform in size; whereas the grid
segments in vertical ($y$-) direction are progressively refined towards the bottom boundary. The
boundary conditions for both pressure field and velocity field are periodic in both $x$- and
$z$-directions.  In $y$-direction, zero-gradient boundary condition is applied for pressure field;
no-slip wall condition is applied at the bottom for velocity field, and slip wall condition is
applied on the top. A pressure gradient is applied on the whole domain to maintain the constant bulk
flow velocity. To provide a bottom boundary condition for the moving particles, three
layers of fixed particles are arranged hexagonally at the bottom~\citep{gupta16cfd}. A
perturbation of the first layer of the fixed particles is added to avoid moving particles from
forming a perfect lattice. This is because the perfect lattice of sediment particles can
significantly reduce the sediment transport rate. The properties of the sediment
particles and fluid flow used in the present simulations are consistent with the numerical
benchmarks~\citep{KidanemariamUhlmann2014}. To be consistent with the numerical benchmark, the
linear contact model is also used. It should be noted that the viscous damping effect
during the particle collision should be considered. Hence, the collision restitution coefficient
used in the present work is 0.3, which is smaller than the coefficient for dry quartz. This is used
in the DEM simulations of sediment transport problems~\citep{Schmeeckle2014,KidanemariamUhlmann2014}.

The initialization of the present numerical simulations follows previous
studies~\citep{KidanemariamUhlmann2014,SunXiao2016b}. A separate simulation of particle settling
without considering the hydrodynamic forces is performed to obtain an initial configuration of the
particles. In this settling simulation, particles fall from random positions under gravity with
inter-particle collisions. To initialize the fluid flow, each simulation first runs 20 flow-through
times with all particles fixed at the bottom.

\section{Results}
\subsection{Shear velocity decomposition and averaged transport conditions}

The simulation conditions listed in Table \ref{tab:param-all} gave rise to the development of mobile
bedforms. Because of this, the pressure gradient needed to maintain the unit width discharge of
water had to increase from the initial value within each simulation as more and more particles began
to move, collide, and self-organize into bedforms. As will be discussed in the next section,
bedforms grew in size from Case 1 to Case 2, and then began to wash out in Cases 3 and 4. As such,
the total boundary resistance increased from Case 1 to 2, decreased from Case 2 to 3, and then
increased again between Cases 3 and 4. This up and down behavior in the total resistance occurred
even though the fluid discharge and velocity monotonically increased with each case or Reynolds
number. 

To make the results of the simulations more comparable to classic sediment transport theory, we
decomposed the total shear stress $\tau$ into a frictional component $\tau'$ and the bedform
and collision-drag component $\tau''$ with $\tau = \tau' + \tau''$.  The frictional component,
$\tau'$, was calculated from the force needed to drive the flow at the given flow rate before
significant particle motion or bedform development. This is the friction between the water and the
bed at the sediment grain size scale.  In contrast, the total shear stress $\tau$ values were
obtained from the average of the total force needed to sustain the unit width discharge once the
bedforms had fully developed. The contributions from the form drag and the collisions between
suspended and the immobile particles were then inferred by taking the difference between the two,
i.e., $\tau'' = \tau - \tau'$.  Finally, following the conventions in sediment transport literature,
we use the corresponding friction velocities defined as $u_* = \sqrt{\tau/\rho_f}$, $u_*' =
\sqrt{\tau'/\rho_f}$ and $u_*'' = \sqrt{\tau''/\rho_f}$, respectively, to denote the three shear
stress components.  Shear velocity values for these three components are given in
Table~\ref{tab:dune-transport} along with key contextual sediment transport parameters, all of which
were calculated using the friction shear stress $u_{*}'$. We have done the partition in a way that
mimics what is often done in sediment transport calculations.  That is, where the skin component of
the drag is related to the drag that would be present given a fixed bed (same grain size but no bed
forms), and form drag is the difference between the total drag (in the presence of bedforms and
grain collisions) and the skin friction based on grain size.

\begin{figure}[htbp] 
  \centering
  \subfloat[Shields diagram]{
    \includegraphics[width=0.45\textwidth]{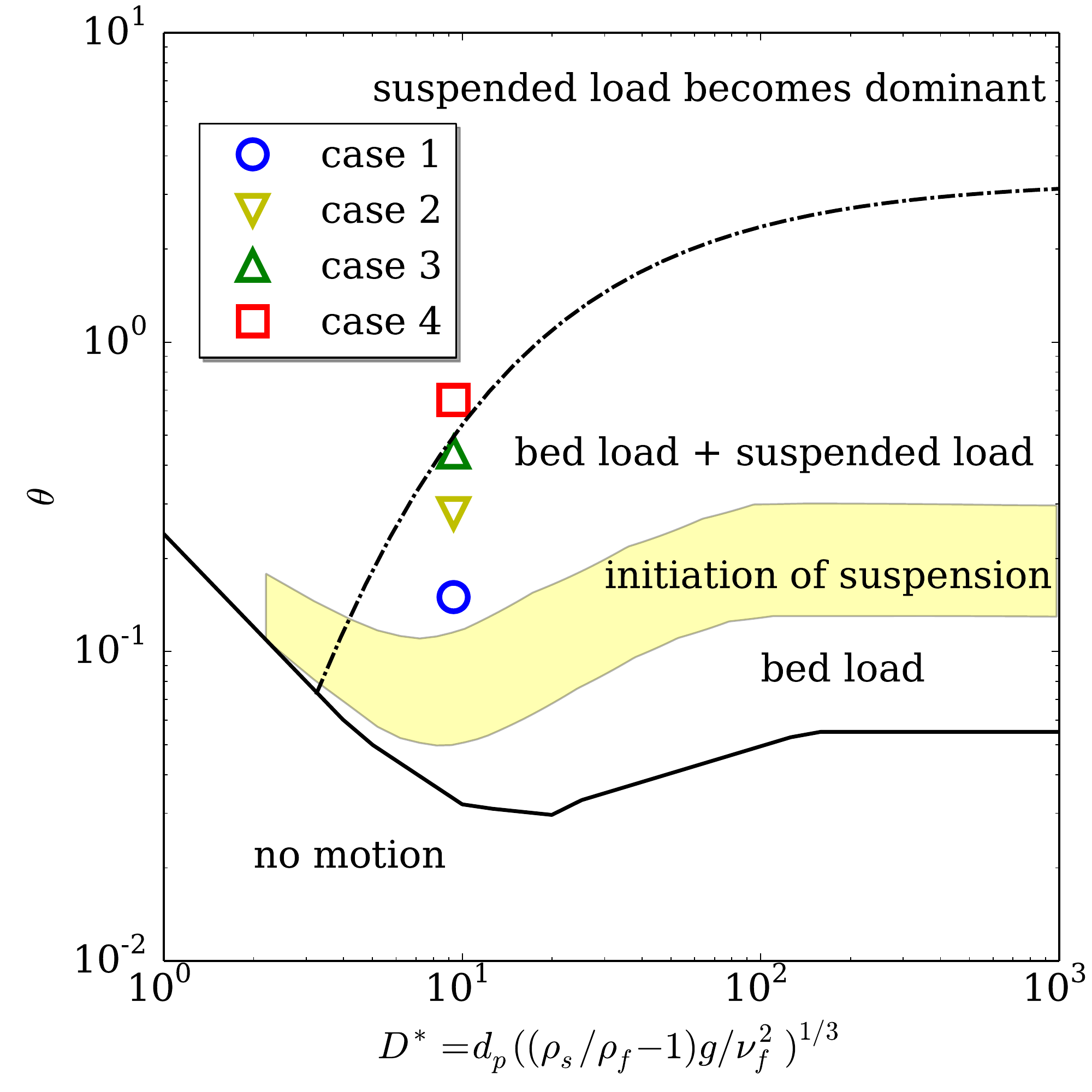}
  }
  \subfloat[phase diagram of dune generation]{
    \includegraphics[width=0.45\textwidth]{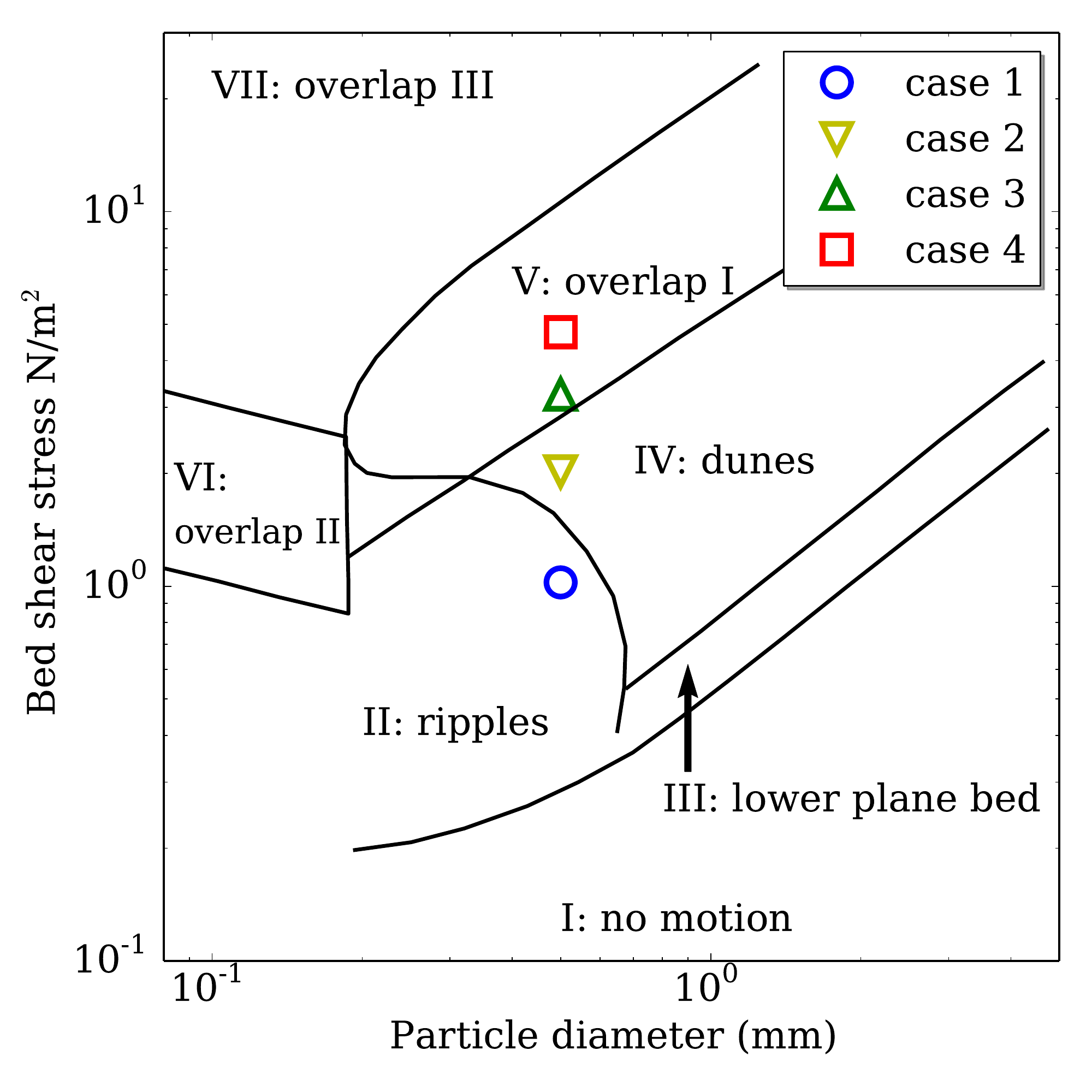}
  }
  \caption{The phase diagrams of different sediment transport conditions in the present study. In
  the phase diagram, overlap I is the overlap region of dune, upper plane bed, and antidunes;
  overlap II is the overlap region of ripples, upper plane bed, and antidunes; overlap III is the
  overlap region of upper plane bed and antidunes. }
  \label{fig:regimes}
\end{figure}

The conditions in the numerical simulations are compared to the regime map obtained from the summary
of experimental results, shown in Fig.~\ref{fig:regimes}. Figure~\ref{fig:regimes}(a) is plotted
according to the Shields diagram of sediment transport. The Shields curve is plotted as solid line
to denote the threshold for incipient sediment motion~\citep{rijn84se1}. The yellow (shaded) region
indicates the criteria for the initiation of suspended load. The dash-dotted curve denotes the
threshold that the suspended load becomes dominant~\citep{bagnold1966,naqshband14ubf}.  It can be
seen in Fig.~\ref{fig:regimes}(a) that in Case 1, the suspended load is negligible. When the flow
rate increases, the suspended load increases and becomes dominant in Case 4. The phase diagram
of the generation of bedforms is plotted in Fig.~\ref{fig:regimes}(b)
according to~\cite{southard90bed}. It can be seen that the conditions for Case 1 fall in region II,
where small ripples are generated in experimental studies. Dune-like patterns are observed in the
experimental studies when the conditions are consistent with Case 2.  For Case 3 and 4, the
hydraulic conditions fall in region V, which is an overlapped region of dunes, upper plane beds, and
antidunes.

\begin{table}[htbp]
  \caption{Resulting shear velocities, sediment transport conditions, and bedform features at
    different bulk Reynolds numbers $\mathrm{Re_b}$. All transport conditions are calculated
    with the skin friction shear velocity, $u_{*}'$.  Bedform characteristics were
    given for $t^* \in [700,750]$. } 
 \begin{center}
 \begin{tabular}{lcccc}
   \hline       & Case 1 & Case 2 & Case 3 & Case 4 \\
   \hline
   Shear velocity decomposition &\\
   \qquad total shear velocity $u_{*}$~[m/s]           & 0.037 & 0.081 & 0.072 & 0.088 \\
   \qquad fluid friction shear velocity $u_{*}'$~[m/s]       & 0.032 & 0.045 & 0.057 & 0.069 \\
   \qquad collisional and form shear velocity  $u_{*}''$~[m/s] 
        & 0.019 & 0.067 & 0.043 & 0.055 \\ 
   \qquad ratio of collisional/form drag to total, ${u_{*}''}^2/u_{*}^2$ 
        & 0.25 & 0.69 & 0.37 & 0.39 \\ 
   Sediment transport conditions &\\
   \qquad particle Reynolds number $\mathrm{Re_p}= d_{p}\sqrt{gR_{s}d_{p}}/\nu$  & \multicolumn{4}{c}{43}\\ 
   \qquad critical Shields parameter $\theta_{cr}$       & \multicolumn{4}{c}{0.032}\\ 
   \qquad $w_s/u_{*}'$                             & 1.88  & 1.33  & 1.05  & 0.87 \\
   \qquad Shields parameter $\theta$            & 0.15  & 0.28  & 0.44  & 0.65 \\ 
   \qquad mobility condition $\theta/\theta_{cr}$  & 4.6  & 8.5  & 13.6  & 20.0 \\ 
   \qquad total load $q_{t}^{*}=q_{t}/d_{p}\sqrt{gR_{s}d_{p}}$		& 0.04 & 1.12 & 0.97 & 4.65 \\
   \qquad suspended load fraction, $\alpha_{s}=q_{s}^{*}/q_{t}^{*}$ & 0.14 & 0.17 & 0.43 & 0.66 \\
   Bedform characteristics &\\
   \qquad wavelength ($L_d/d_p$)               & 160   & 312   & 312 $^{(a)}$   & N.A. \\
   \qquad height ($H_d/d_p$)                   & 3.1   & 11.6  & 7.2  $^{(a)}$   & N.A.\\
   \qquad migration velocity ($U_d/U_b$)           & 0.017 & 0.053 & 0.088  $^{(a)}$ & N.A.\\
   \qquad bedform transport rate $q_{\mathit{bf}}^{*}=q_{\mathit{bf}}/d_{p}\sqrt{gR_{s}d_{p}}$ & 0.06 & 0.93 & N.A. & N.A. \\
   \hline
  \end{tabular}
 \end{center}
Note: (a) While bedform characteristics are presented for Case 3, it should be  noted that any
structure that was present at a given time was highly unstable and  significantly less dune-like
than the bedforms in Cases~1 and~2.
 \label{tab:dune-transport}
\end{table}

\subsection{Bedform morphology and evolution}
\label{sec:ap-bedform}

The bedform morphology and evolution obtained from numerical simulations of various bulk Reynolds
numbers are compared in this section. In our previous study~\citep{SunXiao2016b}, we had already
demonstrated that the LES--DEM model is capable of self-generating bedforms with physically
meaningful heights, wavelengths, and migration velocities. The simulated characteristics had been
validated against experimental measurements and interface-resolved simulations reported in the
literature~\citep{KidanemariamUhlmann2014}. The bedforms that develop in the present study are
similar.  Based on the calculated Shields parameter, $\theta = 0.15$ to 0.65, and a particle
Reynolds number of $\mathrm{Re_p}=d_{p}\sqrt{gR_{s}d_{p}}/\nu=43$ (Table \ref{tab:param-all}), where
$R_s=(\rho_s-\rho_f)/\rho_f$ is the submerged specific gravity.  We can say that the simulation
conditions are inline with those that should give rise to bed load bedforms that transition to
suspension dominated bedforms according to the Shields-Parker river sedimentation diagram
\citep{GarciaEtal2000}.

\begin{figure}[htbp]
  \centering
  \subfloat[Case 1: $\mathrm{Re_b}$ = 6,000]{
  \includegraphics[width=0.45\textwidth]{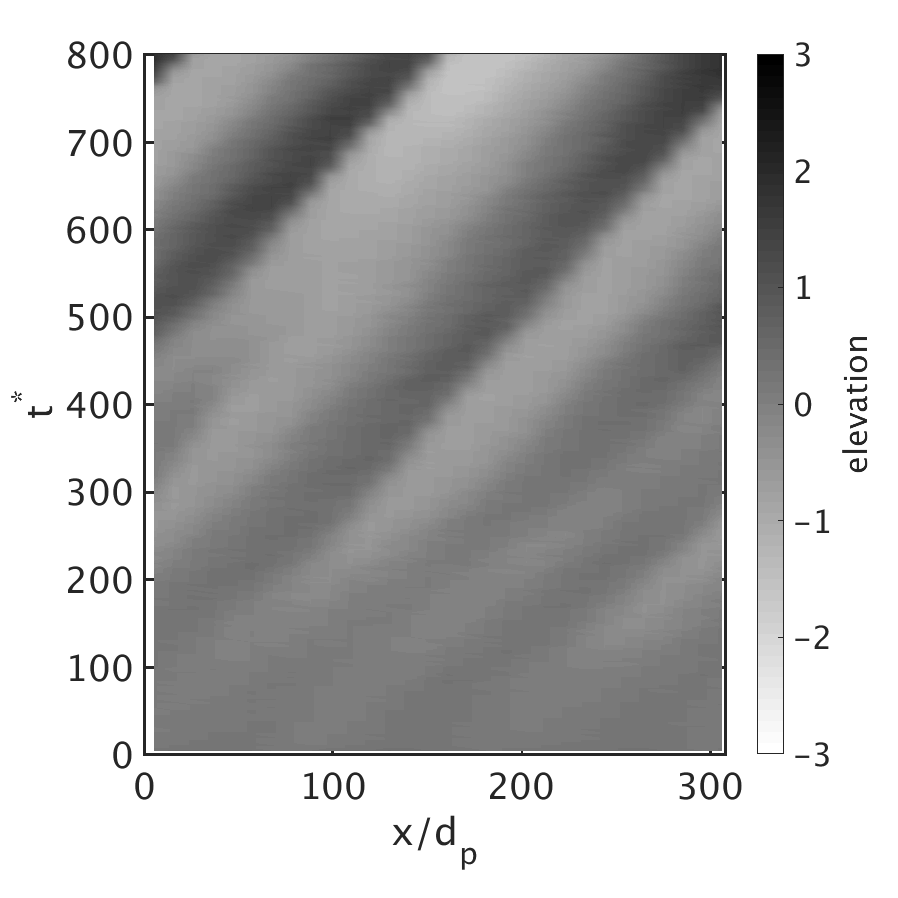}
  }
  \subfloat[Case 2: $\mathrm{Re_b}$ = 8,000]{
  \includegraphics[width=0.45\textwidth]{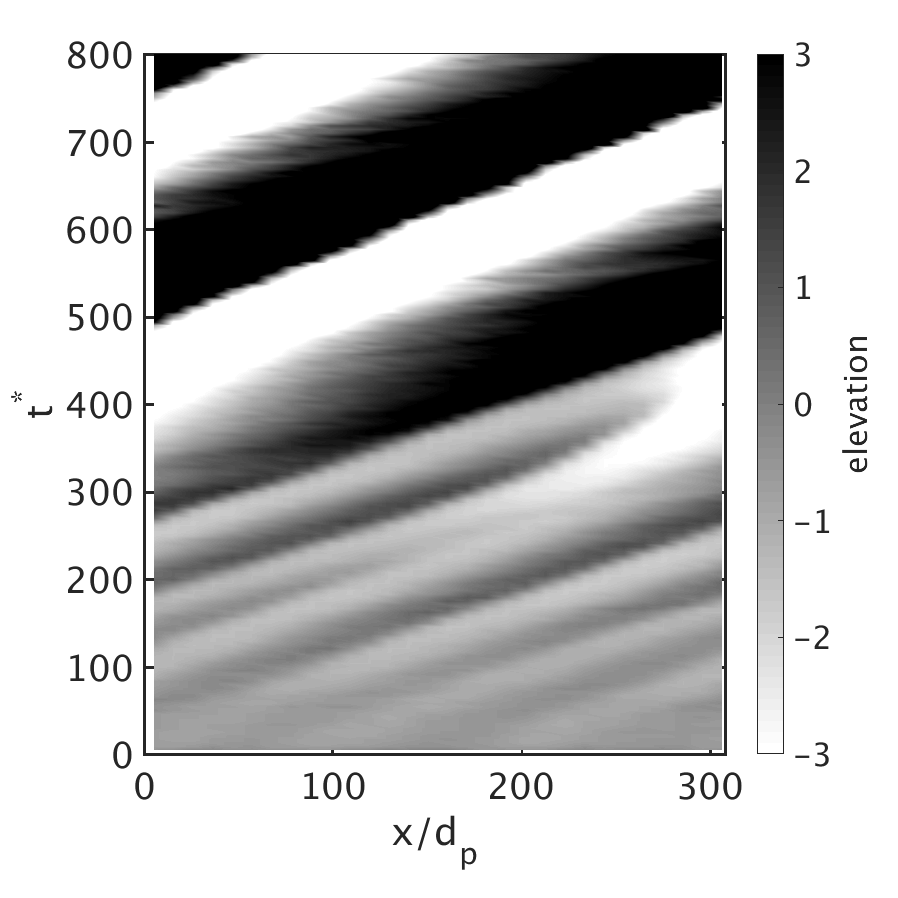}
  }
  \hspace{0.1in}
  \subfloat[Case 3: $\mathrm{Re_b}$ = 10,000]{
  \includegraphics[width=0.45\textwidth]{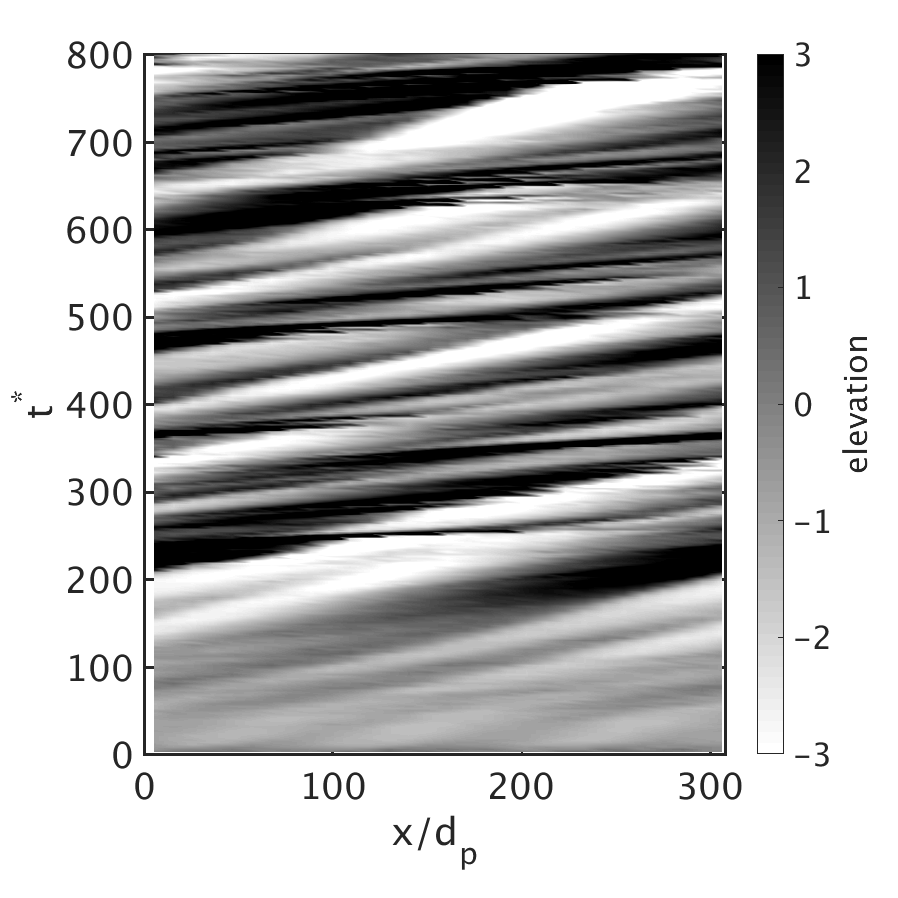}
  }
  \subfloat[Case 4: $\mathrm{Re_b}$ = 12,000]{
  \includegraphics[width=0.45\textwidth]{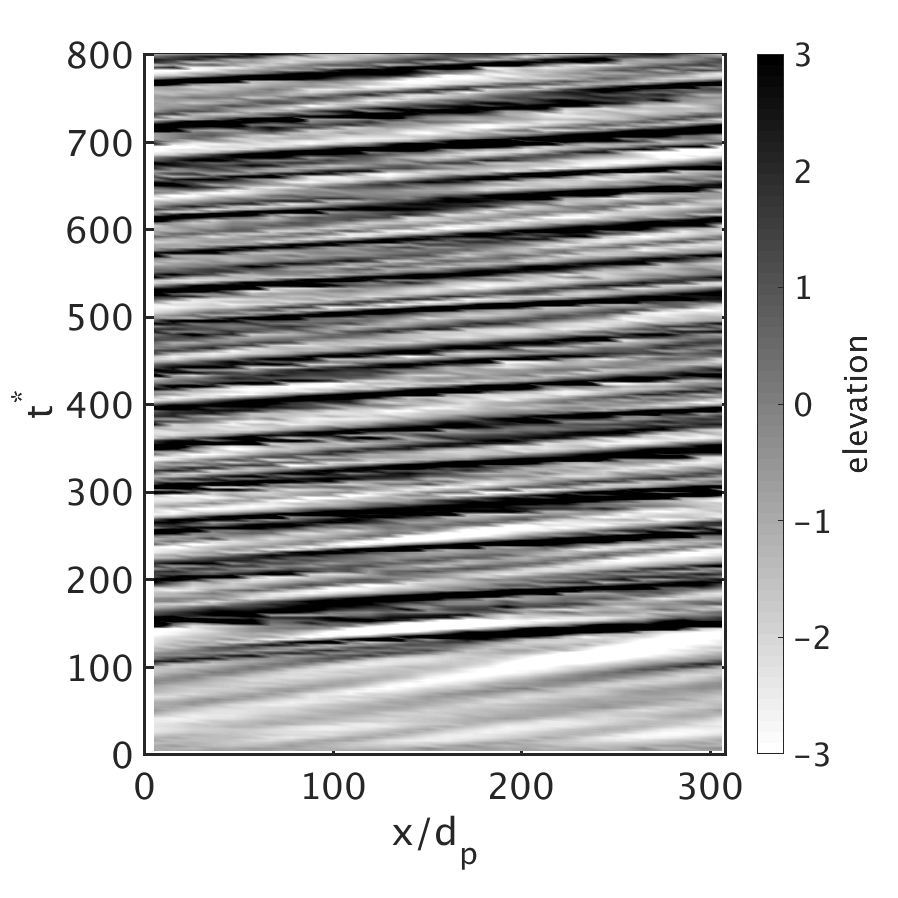}
  }
  \caption{Time-space evolution of the bedforms at bulk Reynolds numbers $\mathrm{Re_b}$ from 6,000 to
    12,000. The color of the contour indicates the height of the bed surface compared to the
    initial flat bed.}
  \label{fig:time-space-re}
\end{figure}

The space-time evolutions of the bed surface at various bulk Reynolds numbers are plotted in gray
scale in Fig.~\ref{fig:time-space-re}. This aims to demonstrate the variation of the surface in the
temporal-spatial domain. It should be noted that bedforms generated in the simulations
are three-dimensional, but the lateral-averaged bed surface profiles are plotted. The surface of the
bed is determined by using a threshold value $\varepsilon_s = 0.1$ according
to~\cite{KidanemariamUhlmann2014}. Dark and light oblique stripes, which are the peaks and troughs
of the bedforms respectively, can be observed at all Reynolds numbers. Moreover, the bedforms at all
bulk Reynolds numbers are developed from the perturbations on the flat bed, which is consistent with
the observations by \cite{KidanemariamUhlmann2014}. To facilitate presentation, the physical time
$t$ is normalized by the constant $T = H_b/U_b$, and thus the resulting non-dimensional time is
denoted as $t^* = t/T$.  Figure~\ref{fig:time-space-re}(a) shows that the slope of the stripes after
non-dimensional time $t^* > 400$ is smaller than that of $t^* \in [200,400]$. The decrease of the
slope indicates the bedform migration velocity decreases due to the growth of the bedform, which is
also observed in previous experimental studies~\citep{ouriemi09sd2}.  Since flow rate $q_f$ is
constant and sediment transport rate does not vary significantly when the bedforms are generated,
larger bedforms move slower ~\citep{paarlberg06mm}.  Comparing the slope of the stripes in each case
(Fig.~\ref{fig:time-space-re}), one can observe that migration velocity increases with the Reynolds
number or shear. This occurs even though the bedforms grow in size from Case~1 to~2, indicating an
overall increase in the sediment transport rate due to the faster moving particles and bedforms.

\begin{figure}[htbp]
  \centering
  \subfloat[Case 1: $\mathrm{Re_b}$ = 6,000]{
  \includegraphics[width=0.45\textwidth]{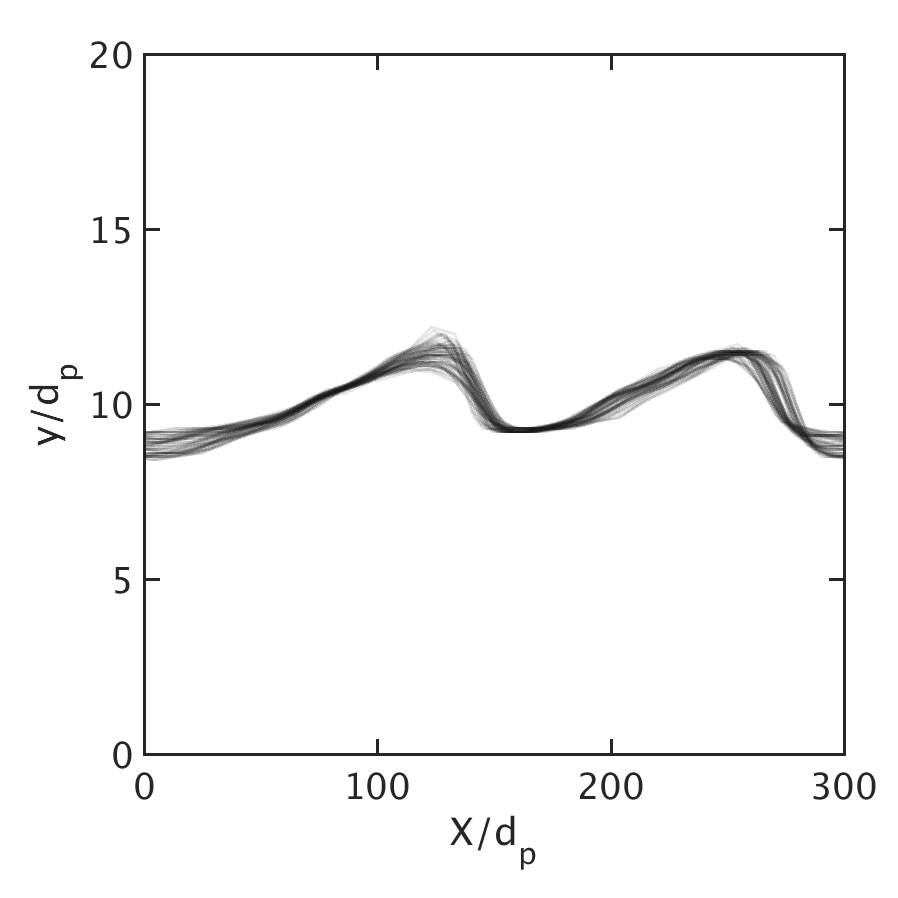}
  }
  \subfloat[Case 2: $\mathrm{Re_b}$ = 8,000]{
  \includegraphics[width=0.45\textwidth]{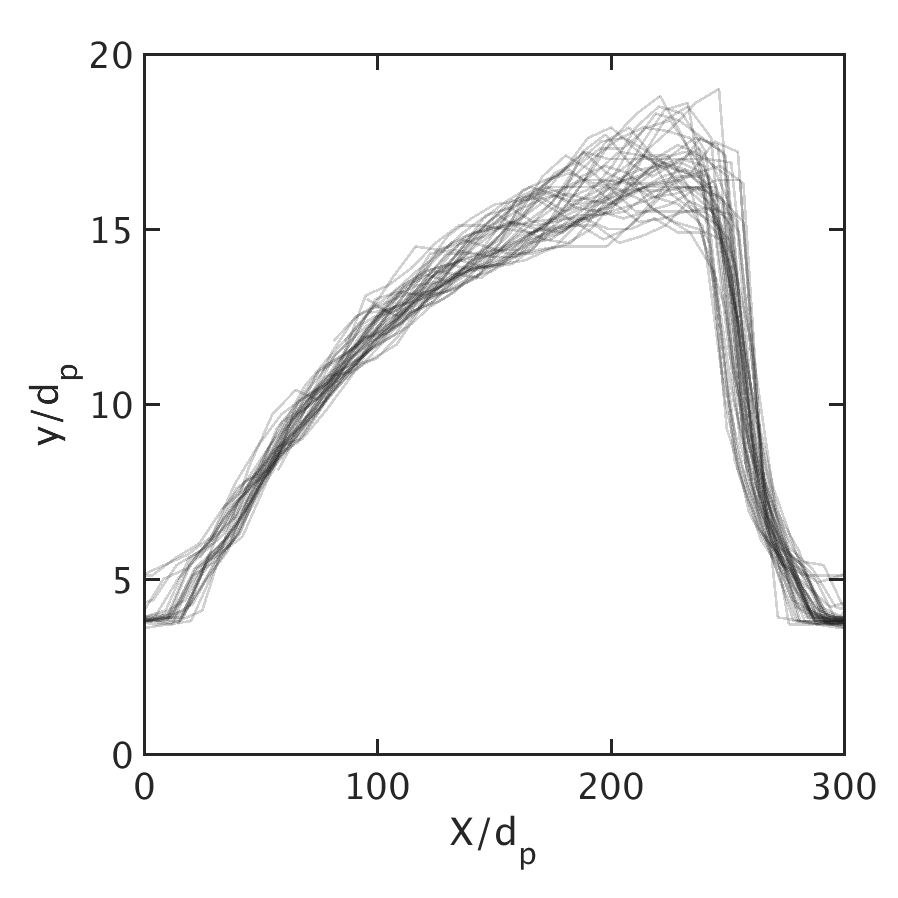}
  }
  \hspace{0.1in}
  \subfloat[Case 3: $\mathrm{Re_b}$ = 10,000]{
  \includegraphics[width=0.45\textwidth]{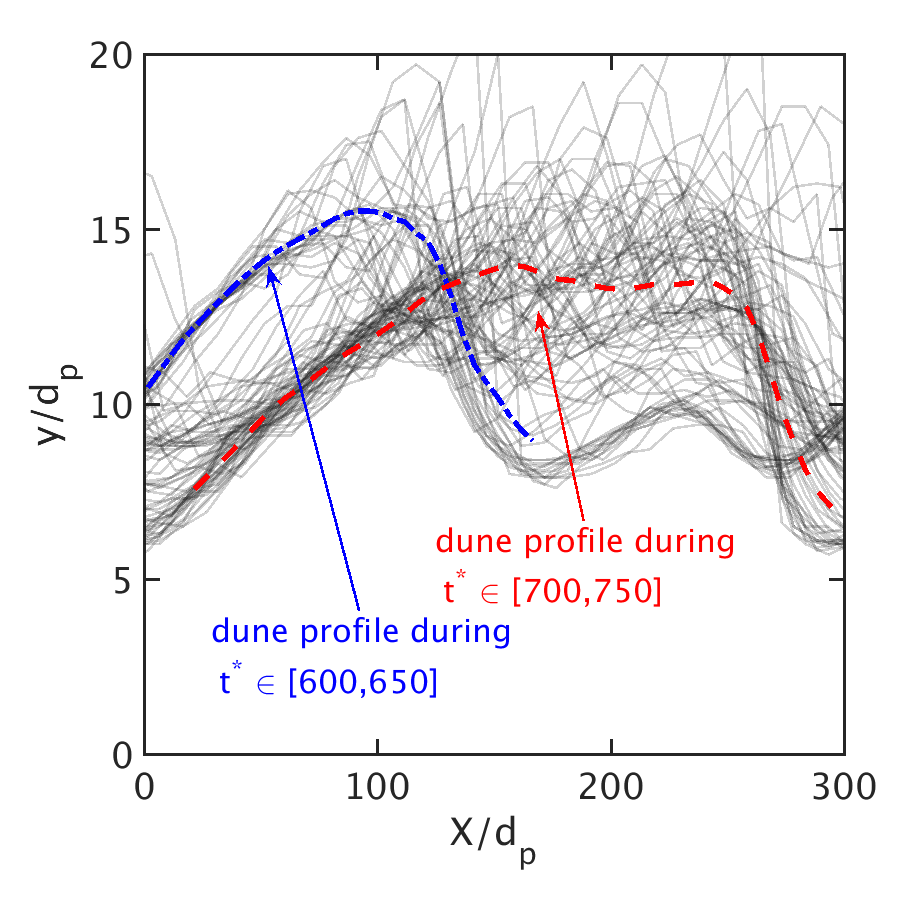}
  }
  \subfloat[Case 4: $\mathrm{Re_b}$ = 12,000]{
  \includegraphics[width=0.45\textwidth]{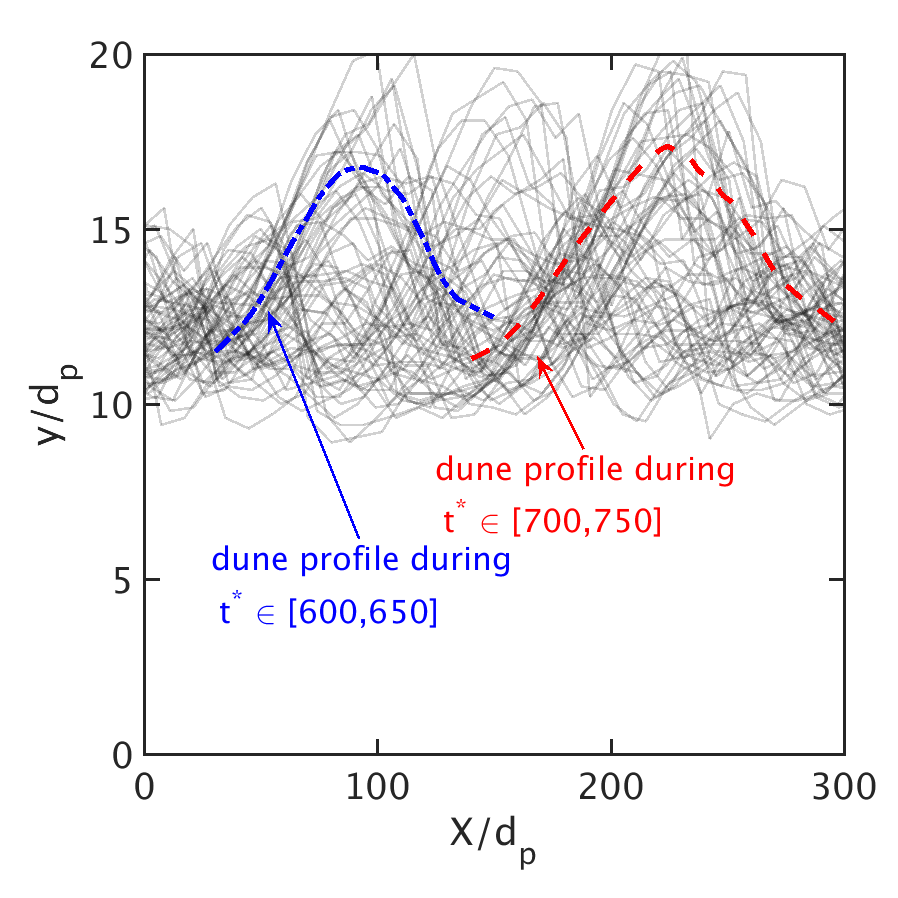}
  }
  \caption{The surface of the bed when $t^* \in [600,800]$ at bulk Reynolds numbers $\mathrm{Re_b}$
    from 6,000 to 12,000, where the non-dimensional time is defined as $t^*$~$=$~$tU_b/H_b$. The
    $x$-axis is the relative longitudinal location with respect to the bedform, $X$, where
    $X$~$=$~$x-U_d t$, $x$ is the fixed downstream direction, and $U_d$ is the migration velocity of
    the bedform.}
  \label{fig:surface-re}
\end{figure}

It can be also seen in Fig.~\ref{fig:time-space-re} that the bedform surfaces are stable in Cases~1
and~2, and unstable in Case~3 and~4. To further investigate the stability of the bedform, the profiles
of the bed surfaces obtained in each case during $t^* \in [600,800]$ are plotted in
Fig.~\ref{fig:surface-re}. The time interval $t^* \in [600,800]$ is selected because
the bedforms are fully developed during this time. The profiles of the bed are plotted using the
relative longitudinal location with respect to the bedform, $X$, where $X=x-U_d t$, $x$ is the fixed
downstream direction, and $U_d$ is the bedform migration velocity.  When the bedform is stable,
bedform surface profiles are overlapping when plotted in the moving-frame.  It can be seen in
Figs.~\ref{fig:surface-re}(a) and \ref{fig:surface-re}(b) that the self-generated bedforms in Case 1
and 2 are stable and the bed surface profiles are highly overlapped. In contrast, the bed profiles
in Fig.~\ref{fig:surface-re}(c) are not overlapping, and there is an offset of about $50d_p$ between
the time-averaged bed profiles obtained when $t^* \in [600,650]$ and $t^* \in [700,750]$. The
offset between the bed profiles at different time intervals shows that the bed surface profile at
bulk Reynolds number $\mathrm{Re_b}$ = 10,000 (Case 3) is less stable than those at lower Reynolds
numbers or shear. This suggests that the present simulation at $\mathrm{Re_b}$ = 10,000 captures the
transition from `dune migration' to `suspended sediment transport'. At even higher bulk Reynolds
number $\mathrm{Re_b}$ = 12,000, shown in Fig.~\ref{fig:surface-re}(d), the time-averaged bed
profiles when $t^* \in [600,650]$ and $t^* \in [700,750]$ fluctuate even more than those at
$\mathrm{Re_b}$ = 10,000. Although bedforms are still generated at $\mathrm{Re_b}$ = 12,000, they are
very unstable because the shear stress of the fluid flow is high enough to wash out the bedforms almost
immediately after they appear. The transition from bedform inception to washout conditions is also
observed in the experimental studies~\citep{bridge88flow,bennett98fluid,baas94flume}. From the
experimental studies, the wavelength of the bedform decrease slightly compared to the
fully-developed bedform, which is consistent with our simulation results. The height of the bedform
decreases more significantly than the wavelength in both experimental measurements and the present
simulations.

To quantitatively compare the features of the bedforms at different bulk Reynolds numbers, the
height $H_d$, the wavelength $L_d$, and the migration velocity $U_d$ of the bedforms obtained in the
present simulations are summarized in Table~\ref{tab:dune-transport}. The height and wavelength of
the bedforms are calculated from the averaged bed surface profiles, and the migration velocity is
obtained from the slope of the stripes in Fig.~\ref{fig:surface-re}. Since the bedforms generated at
$\mathrm{Re_b} \ge 10,000$ are highly unstable, these features are presented during $t^* \in
[700,750]$ when the bedforms appear at all Reynolds numbers. When bulk Reynolds number increases
from 6,000 to 8,000, the wavelength of the bedforms increases from 160$d_p$ to 312$d_p$, and the
height grows from $3.1d_p$ to $11.6d_p$.  This increase of the wavelength and height indicates the
bedforms grows faster at higher Reynolds numbers, which is also observed in the
experiments~\citep{ouriemi09sd2}. This is attributed to the fact that the sediment transport rate
increases significantly when the flow rate increases. As can be seen in
Table~\ref{tab:dune-transport}, the bedform height decreases when bulk Reynolds number increases
from 8,000 to 12,000. Although the sediment transport rate still increases, particles begin to have
longer jump length or move in suspension, resulting in a move towards the washout of the bedforms.
In addition, the increase of the non-dimensional bedform migration velocity is observed in the
present simulations.  When the flow velocity increases, the growth of sediment transport rate is
much larger than the increase of flow velocity, and thus the non-dimensional bedform migration
increases.

\subsection{Individual particle transport characteristics}
\label{sec:ap-individual}
A unique advantage of the LES--DEM approach over low-fidelity, continuum-scale simulation approaches
is its capability to capture the trajectories of individual sediment particles. To fully utilize
this advantage offered by the high-fidelity approach, we compare the trajectories of sediment
particles when moving over the bed surface at different bulk Reynolds numbers. Our aim in plotting
the trajectories of individual sediment particles is to demonstrate the variation of the particle
behavior at different bulk Reynolds numbers and under different sediment transport regimes (i.e.,
bed load, mixed, and suspended load). In addition, the motion of the buried individual particles at
bulk Reynolds number $\mathrm{Re_b}$ = 6,000 is presented. The study of the burial of the individual
particles aims to investigate the trajectories of sediment particles during bedform migration
process after they jumped over the crest on the bedform.

\begin{figure}[htbp] 
  \centering 
  \subfloat[$t_0^*$]{
    \includegraphics[width=0.45\textwidth]{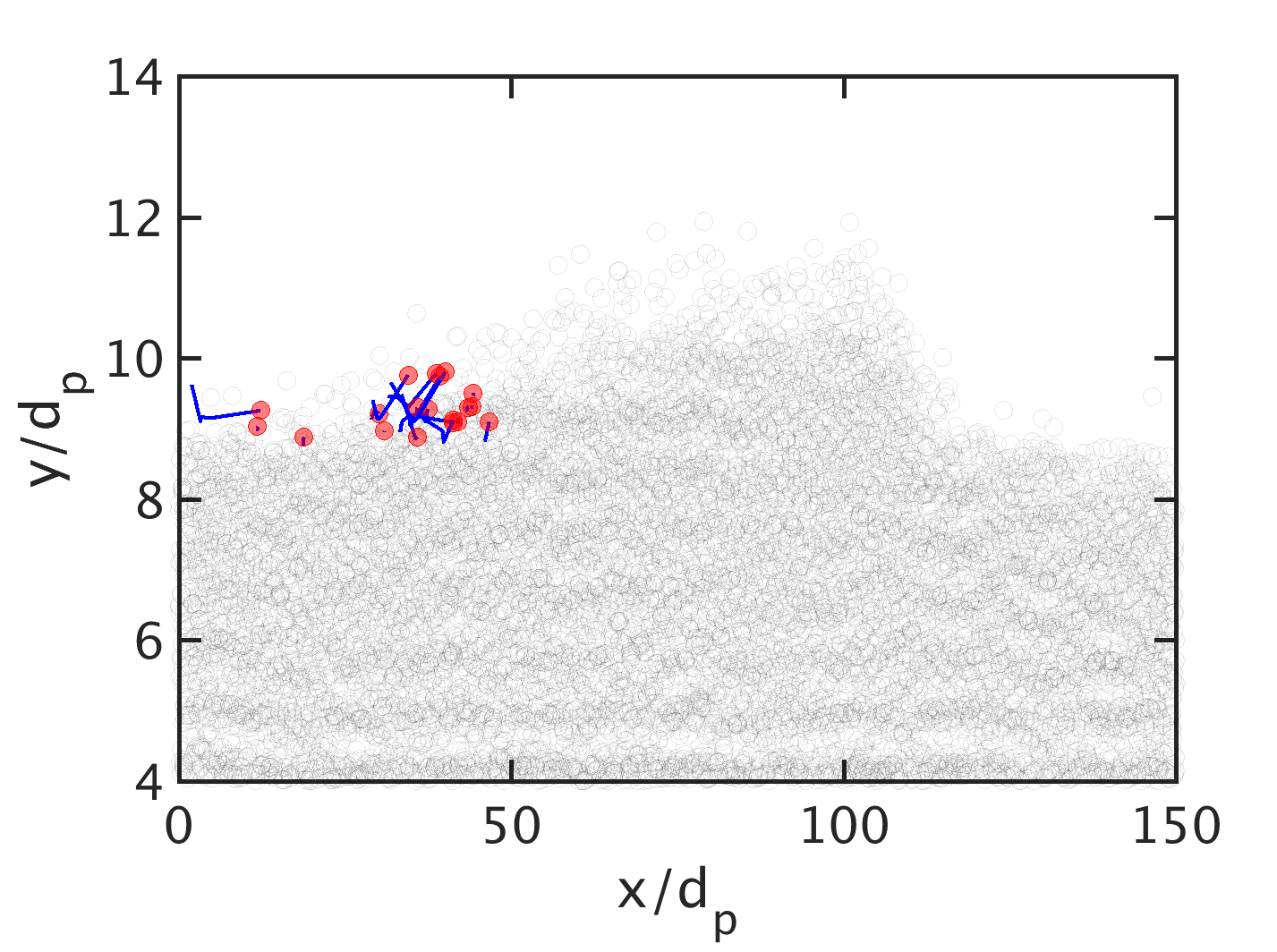} 
  } 
  \subfloat[$t_0^* + 6$]{
    \includegraphics[width=0.45\textwidth]{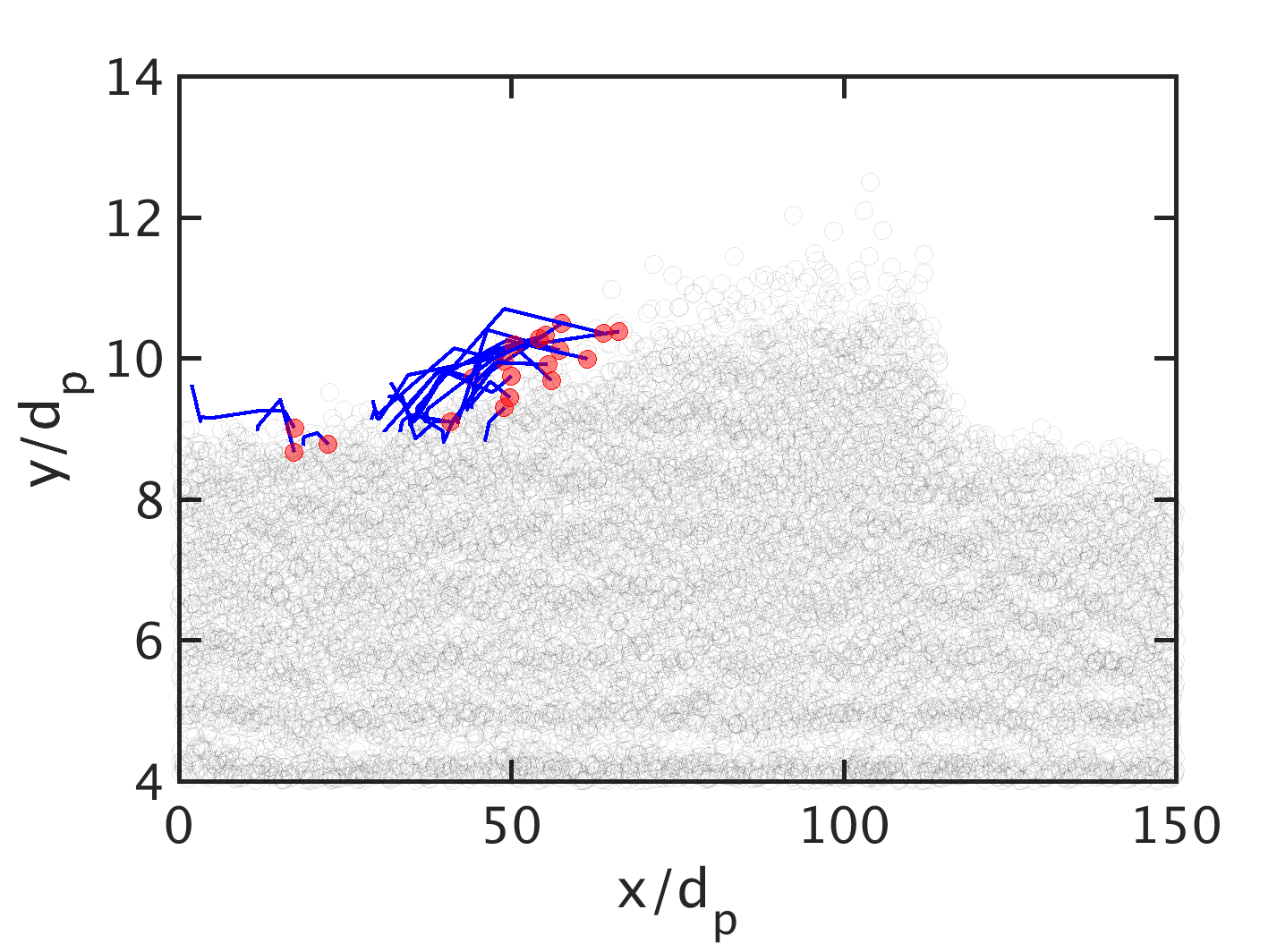}
  } \vspace{0.1 in} 
  \subfloat[$t_0^* + 12$]{
    \includegraphics[width=0.45\textwidth]{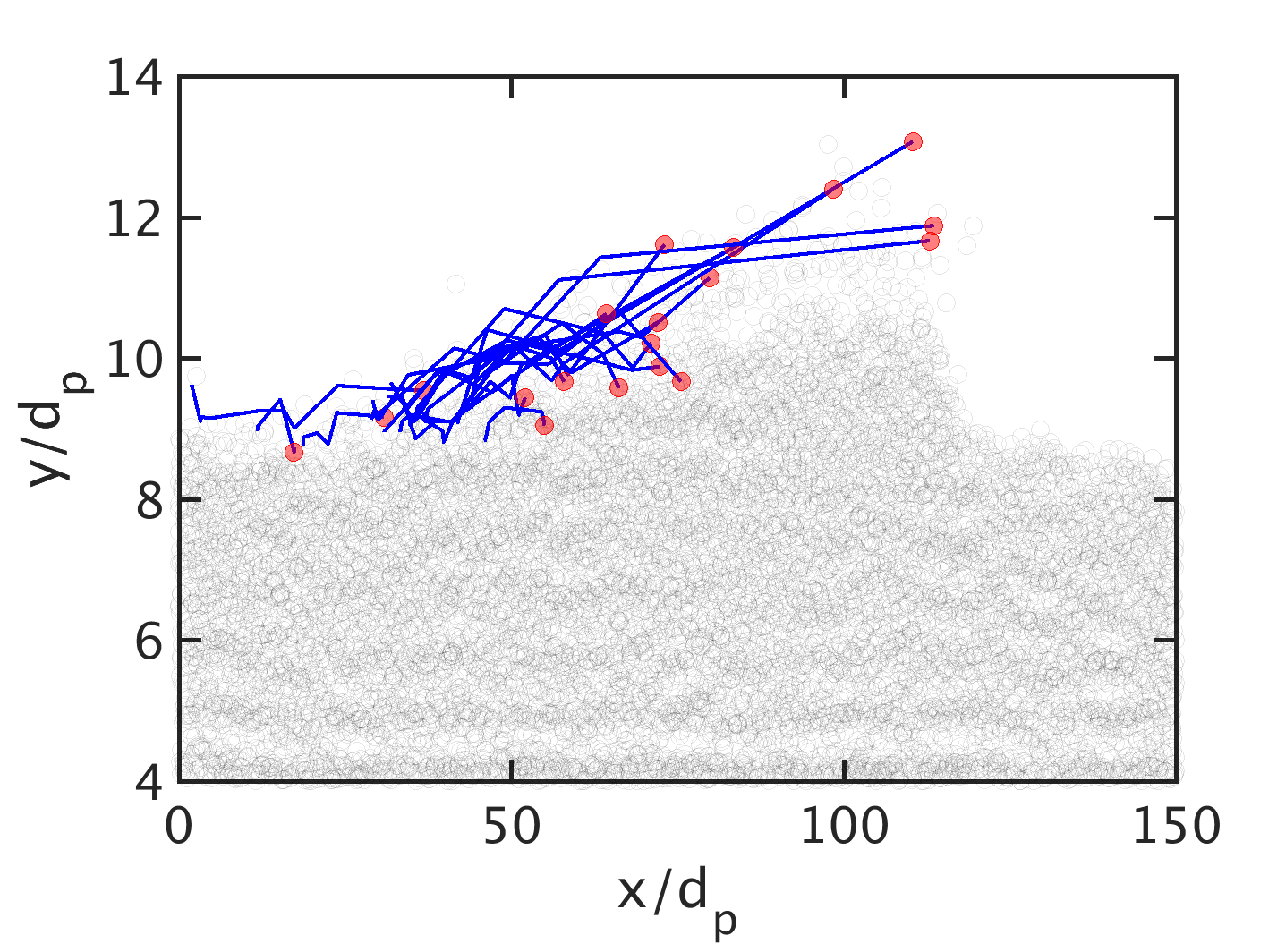}
  } 
  \subfloat[$t_0^* + 18$]{
    \includegraphics[width=0.45\textwidth]{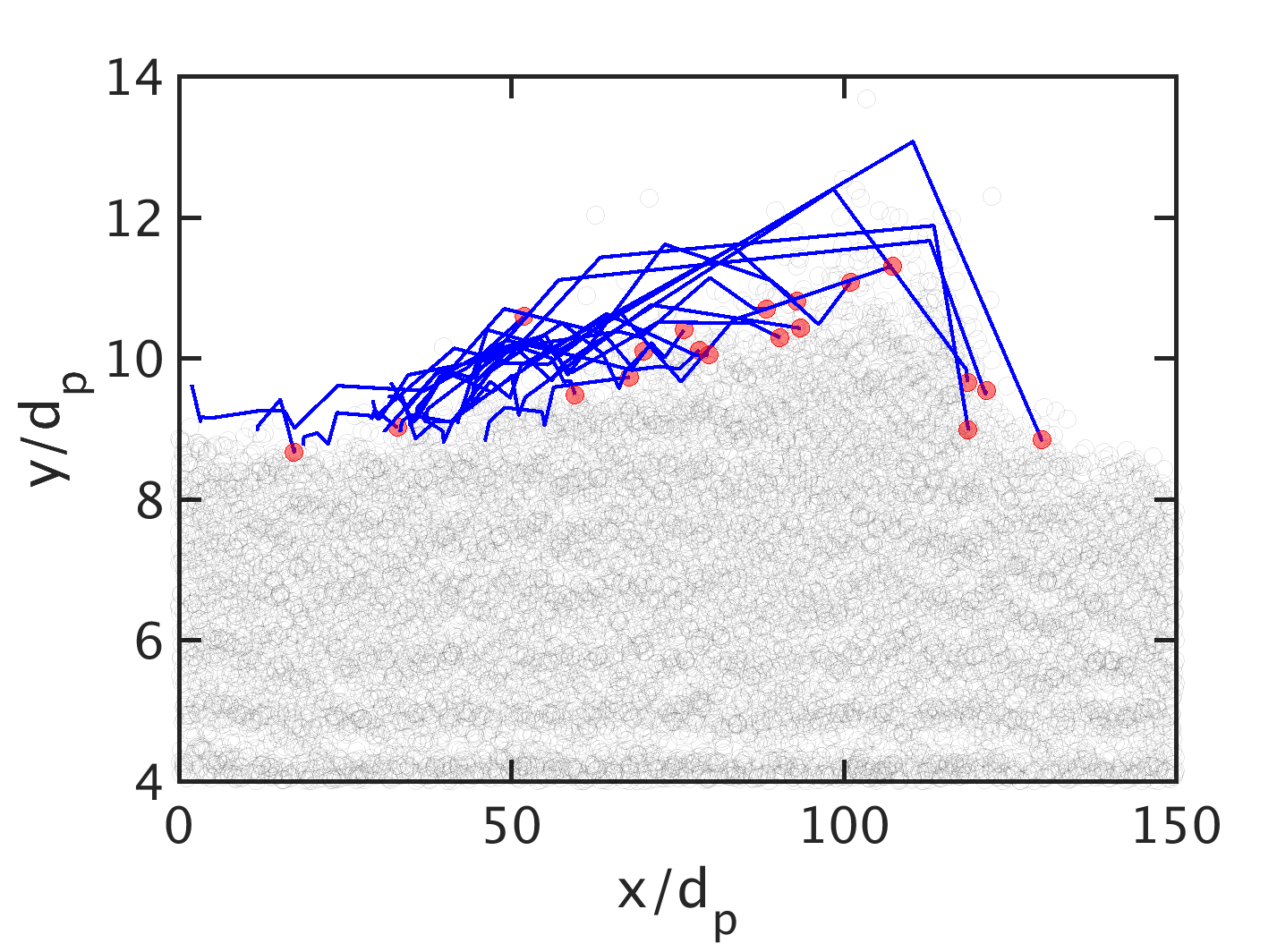}
  } 
  \caption{The trajectories of the sediment particles on the crest of the bedform at bulk Reynolds
    number $\mathrm{Re_b}$~=~6,000 starting from non-dimensional time $t_0^*$~=~700.  The time is
    normalized based on flow depth $H_b$ and bulk velocity $U_b$ as $t^*$~$=$~$t U_b/H_b$.  Each
    circle represents a sphere particle: red circles are highlighted particles; gray particles are
    non-highlighted.  Solid lines are the trajectories of the highlighted particles.}
  \label{fig:trace-re-6000}
\end{figure}

The snapshots of the individual particles jumping over the crest at bulk Reynolds number
$\mathrm{Re_b}$ = 6,000 during $t^* \in [700, 718]$ are shown in Fig.~\ref{fig:trace-re-6000}. To
demonstrate the saltation process of the sediment particles, we randomly selected 20 particles with
initial velocity $u_{p} > 0.1$~m/s on the stoss side and plotted their trajectories. The interval
between two consecutive snapshots is $\Delta t^* = 6$, or equivalently $\Delta t = 0.2$~s at
$\mathrm{Re_b} = 6,000$. It can be seen in the figures that overall behavior of the selected
particles on the bed surface is similar. On the stoss side, these selected particles gain energy
from the fluid flow and move faster than the bedform. However, on the lee side, the flow velocity
decreases after the peak of the bedform, and thus the deposition of the particles can be observed.
Since the particles on the stoss side jump over the peak of the bedform and deposit on the lee side,
the bedform progresses downstream.

\begin{figure}[htbp] 
  \centering
  \subfloat[$\mathrm{Re_b} = 6,000$]{
    \includegraphics[width=0.45\textwidth]{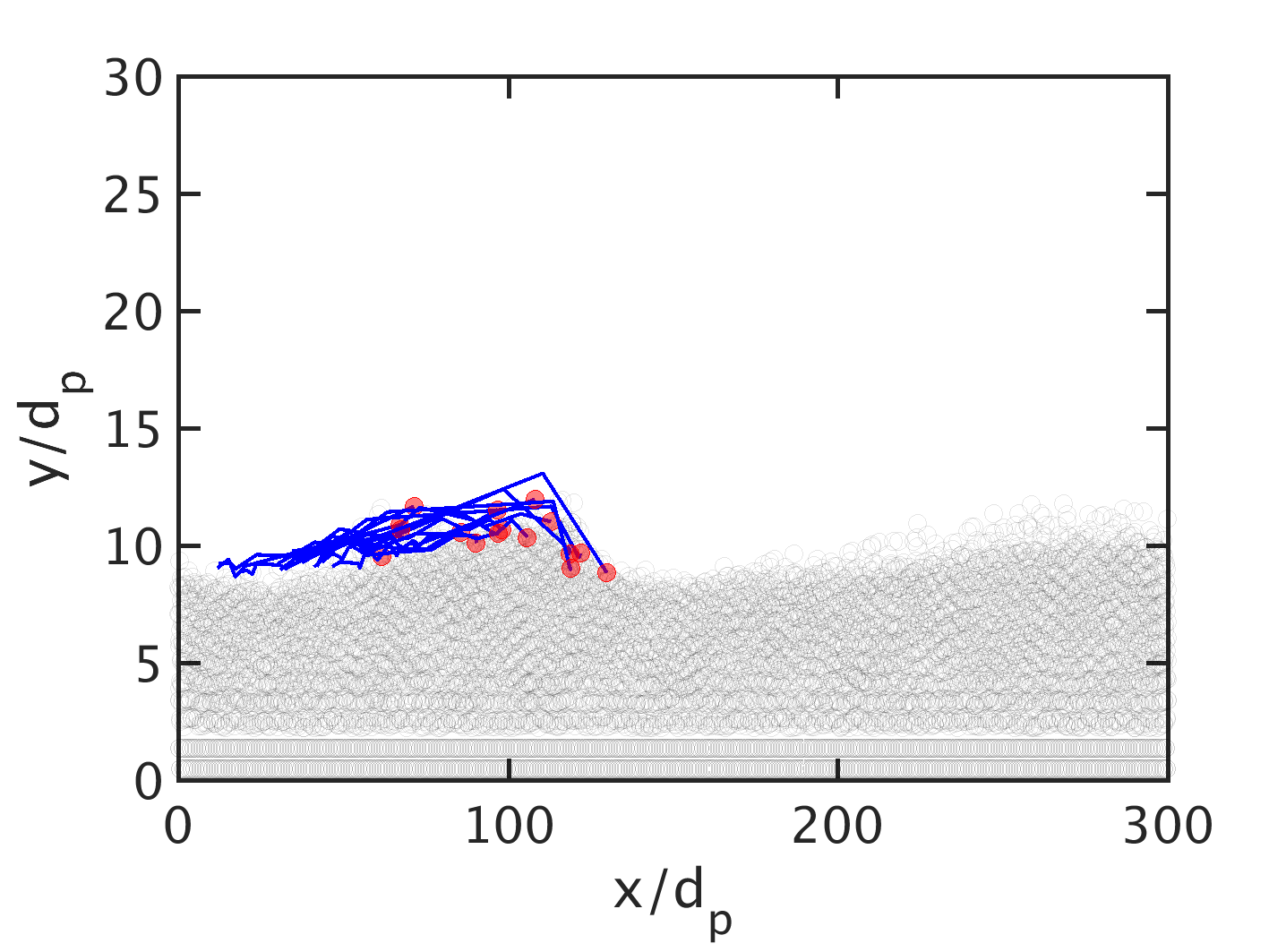}
  }
  \subfloat[$\mathrm{Re_b} = 8,000$]{
    \includegraphics[width=0.45\textwidth]{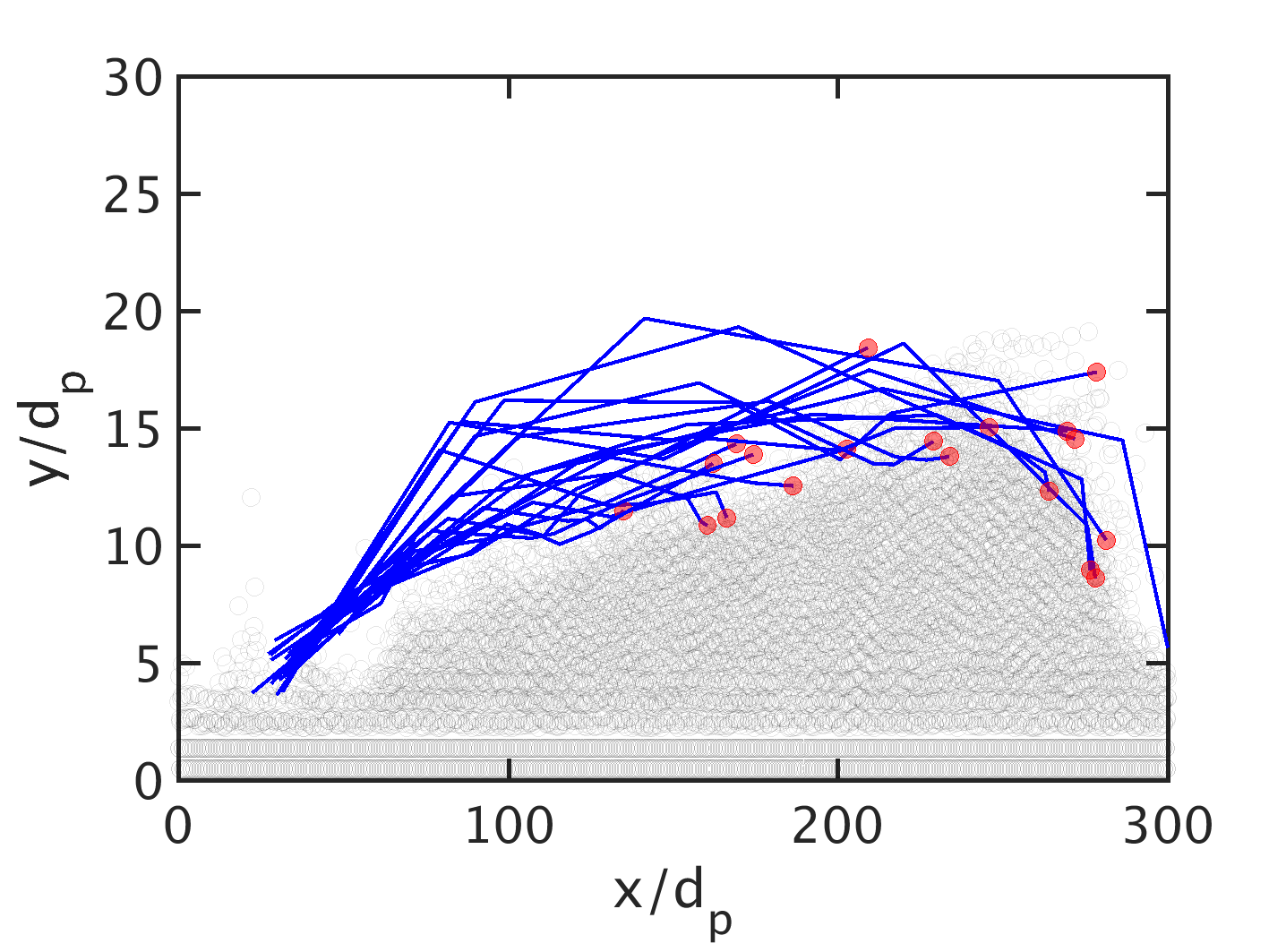}
  }
  \vspace{0.1 in}
  \subfloat[$\mathrm{Re_b} = 10,000$]{
    \includegraphics[width=0.45\textwidth]{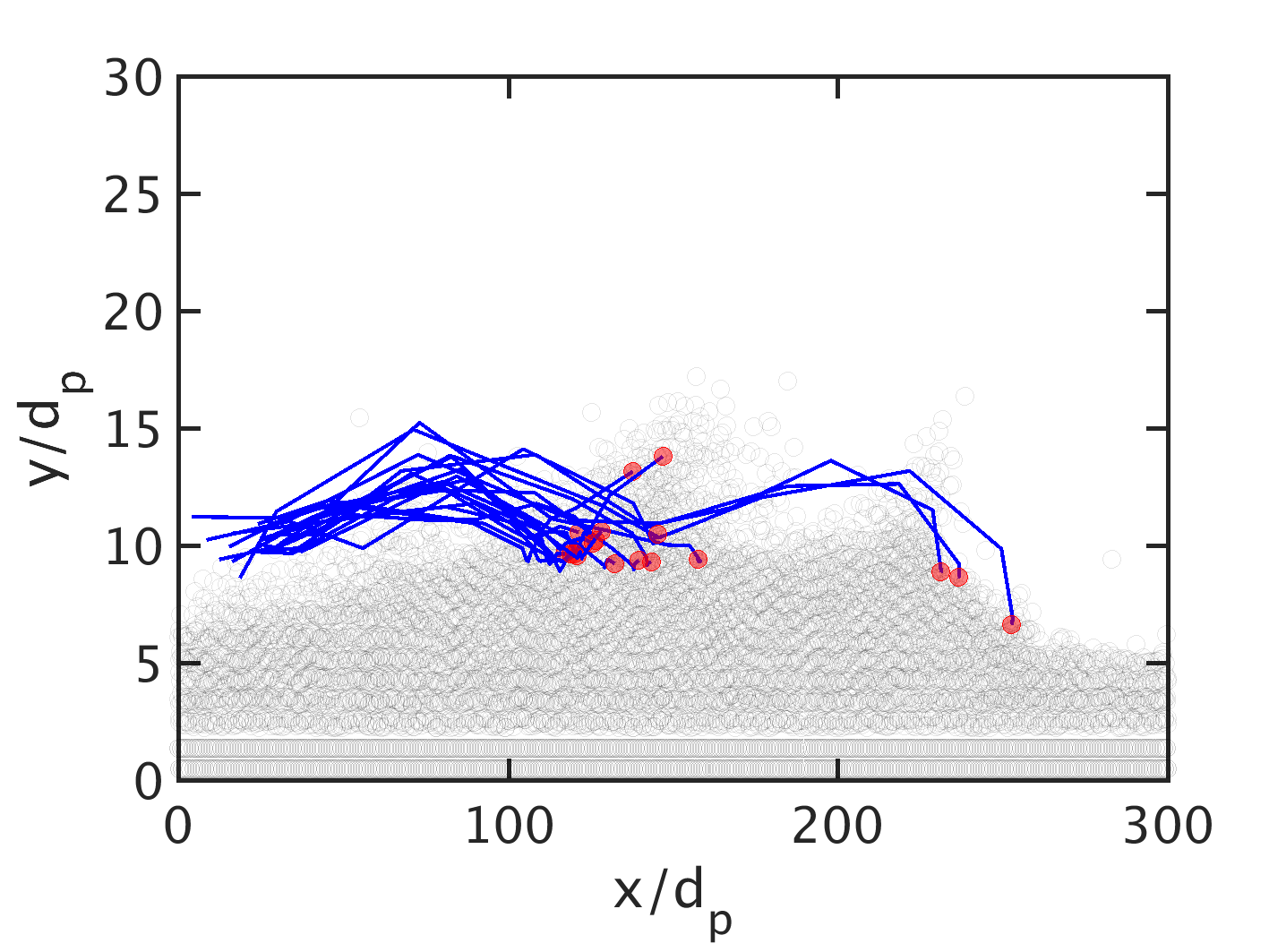}
  }
  \subfloat[$\mathrm{Re_b} = 12,000$]{
    \includegraphics[width=0.45\textwidth]{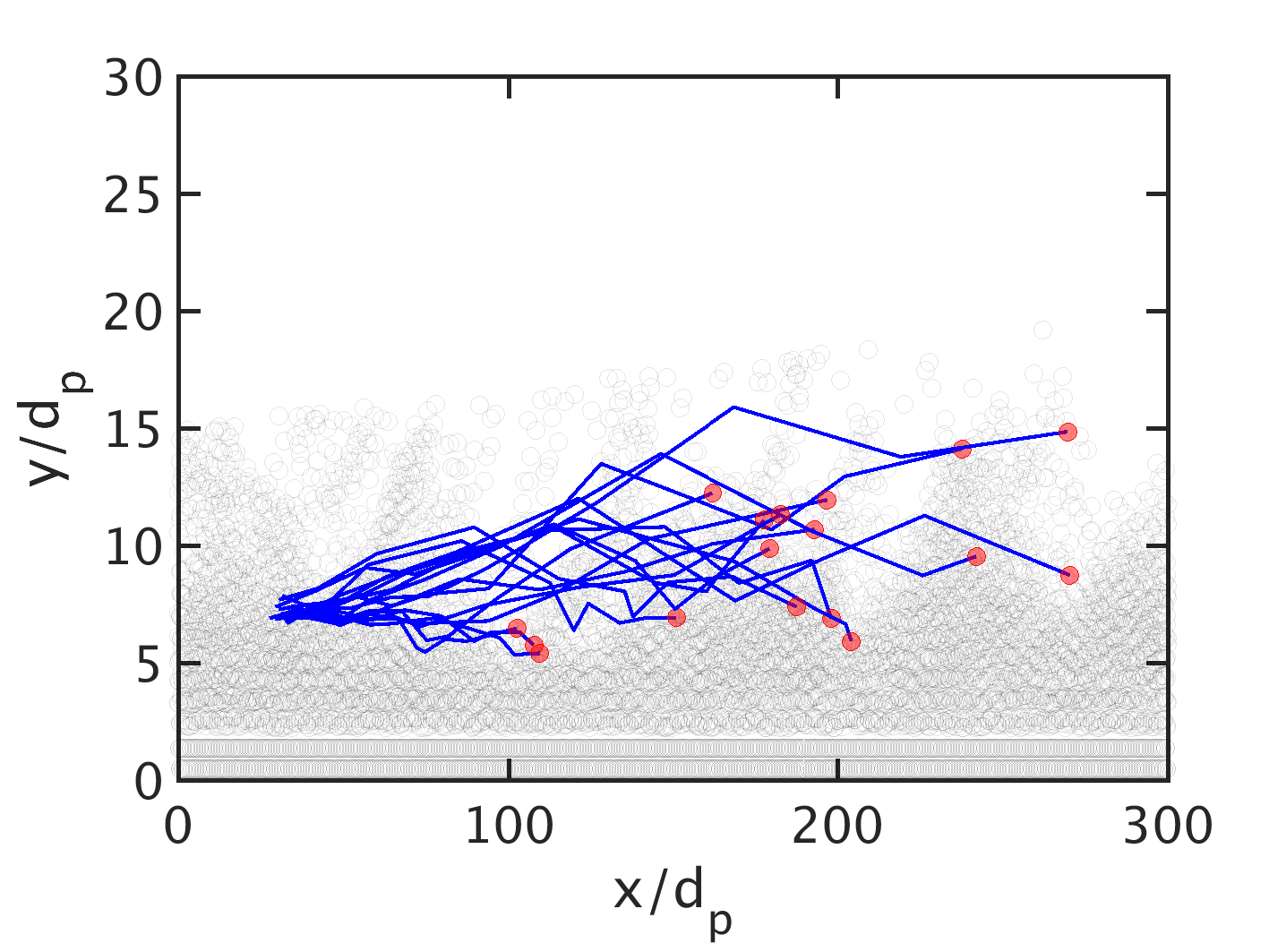}
  }
  \caption{Comparison of the trajectories of the sediment particles on the crest of the bedform during
  $t^*$~$\in$~$[700,725]$ at different bulk Reynolds numbers.  Each circle represents a sphere particle:
  red circles are highlighted particles; gray particles are non-highlighted.  Solid lines are the
  trajectories of the highlighted particles.}
  \label{fig:trace-re-comp-ini}
\end{figure}

The trajectories of individual particles at various bulk Reynolds numbers during $t^* = [700, 725]$
in the longitudinal saltation process are shown in Fig.~\ref{fig:trace-re-comp-ini}, which is to
demonstrate the variation of longitudinal saltation in different flow regimes.  When the bulk Reynolds
number increases from 6,000 to 8,000, the jumping length of the particles significantly increases.
At $\mathrm{Re_b}$ = 8,000, both the bulk velocity and the height of the bedform in the channel are
larger than those at $\mathrm{Re_b}$ = 6,000, and thus the flow velocity on the stoss side of the
bedform increases.  Therefore, the longitudinal jumping velocity of the individual particle in the flow
and the jumping length also grow. However, at bulk Reynolds number $\mathrm{Re_b}$ = 12,000, the
bedform is washed out and behavior of the selected particles is not associated with the surface of the
bed.

\begin{figure}[htbp]
  \centering
  \includegraphics[width=0.6\textwidth]{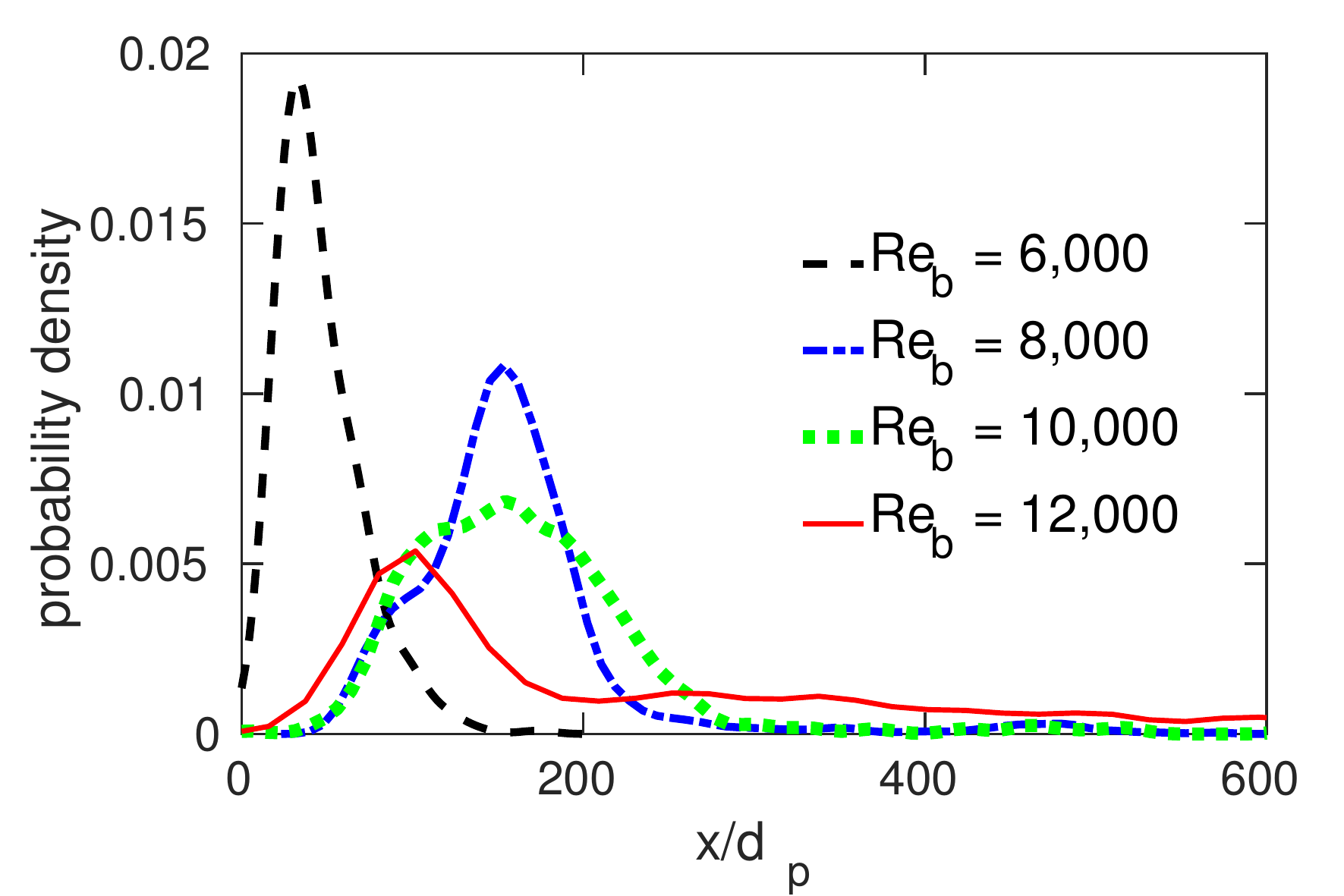}
  \caption{Comparison of the probability density functions of the longitudinal jumping length of the
  sediment particles during $t^*$~$\in$~$[700, 725]$.}
  \label{fig:jump-ini}
\end{figure}

The probability density functions of the particle jumping distances are plotted in
Fig.~\ref{fig:jump-ini}. The saltation of sediment particles is determined by using $U_d$ as threshold
velocity. At bulk Reynolds number $\mathrm{Re_b}$ = 6,000, the peak of the probability density
function is at $x/d_p = 30$, which is much smaller than the bedform wavelength $x/d_p = 150$.  This is
because at $\mathrm{Re_b}$ = 6,000 the longitudinal saltation occurs when the particles are close to
the peak of the bedform. At this Reynolds number, the probability density of the particles jumping over
$x/d_p = 150$ is small. This indicates the number of bypassed sediment particles is negligible and
the wavelength $x/d_p = 150$ in this case is stable. At $\mathrm{Re_b}$ = 8,000, the peak of the
probability density function moves to $x/d_p = 150$. When bedform is generated in the channel, the flow
field on the stoss side would be influenced by the bedform and a vertical flow velocity can be
observed~\citep{NaqshbandEtal2014}. Hence, the resident time of jumping particles is longer and the
peak in the probability density function increases. At $\mathrm{Re_b}$ = 10,000, the peak of the
probability density function is also at $x/d_p = 150$. Due to the increase of bulk flow velocity,
the number of particles jumping over longer distances increases. 

When the Reynolds number continues to increase to 12,000, the probability density of the jumping
distance of $x/d_p > 300$ is significantly larger than those at smaller Reynolds numbers. The
increase of the probability density of long-jumping particles is because the suspended load at
$\mathrm{Re_b}$ = 12,000 is dominant and the number of bypassed sediment is much larger. In
addition, a small peak at $x/d_p = 100$ is observed from the probability density function. The
magnitude of this peak is smaller for $\mathrm{Re_b}$ = 12,000 than for $\mathrm{Re_b}$ = 8,000 and
10,000. This is because the bedform formed at $\mathrm{Re_b}$ = 12,000 is washed out and have
smaller wavelength (shown in Fig.~\ref{fig:surface-re}).

\begin{figure}[htbp]
  \centering
  \subfloat[$t_0^*$]{
  \includegraphics[width=0.45\textwidth]{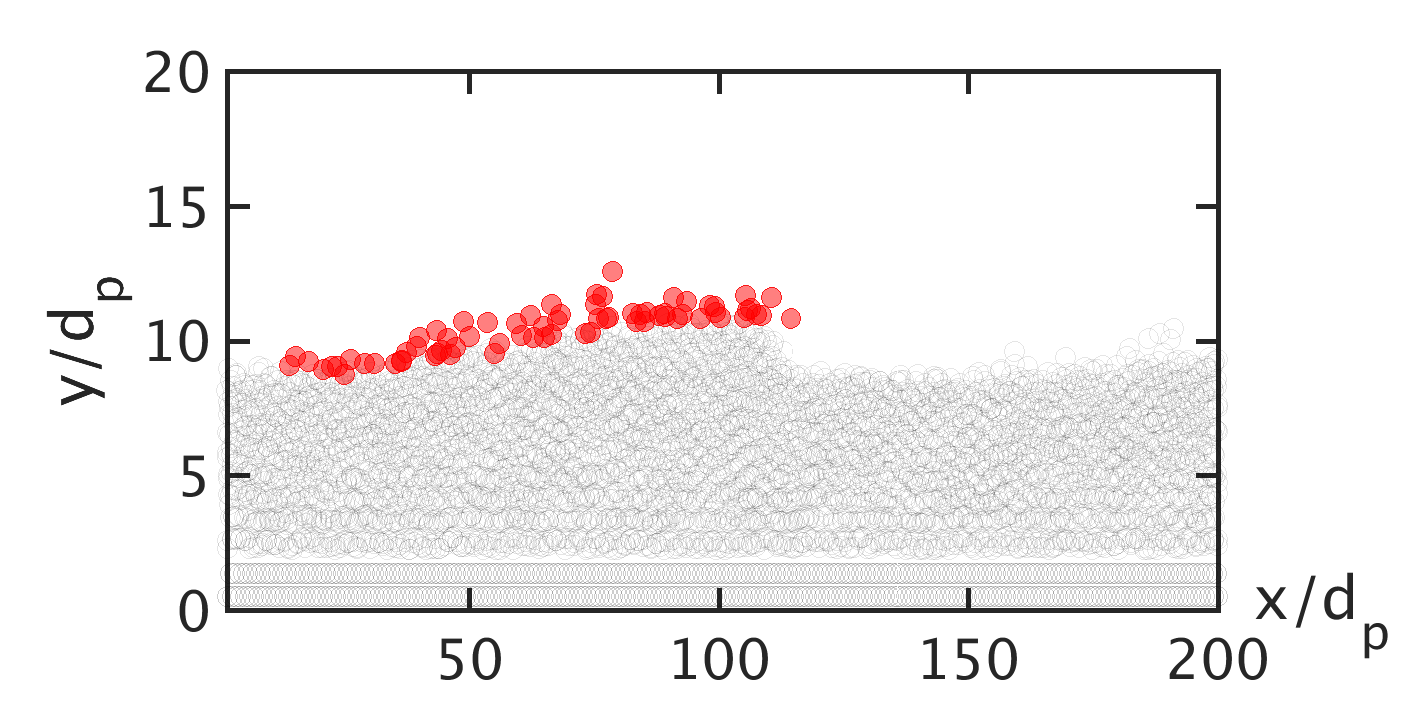}
  }
  \subfloat[$t_0^* + 40$]{
  \includegraphics[width=0.45\textwidth]{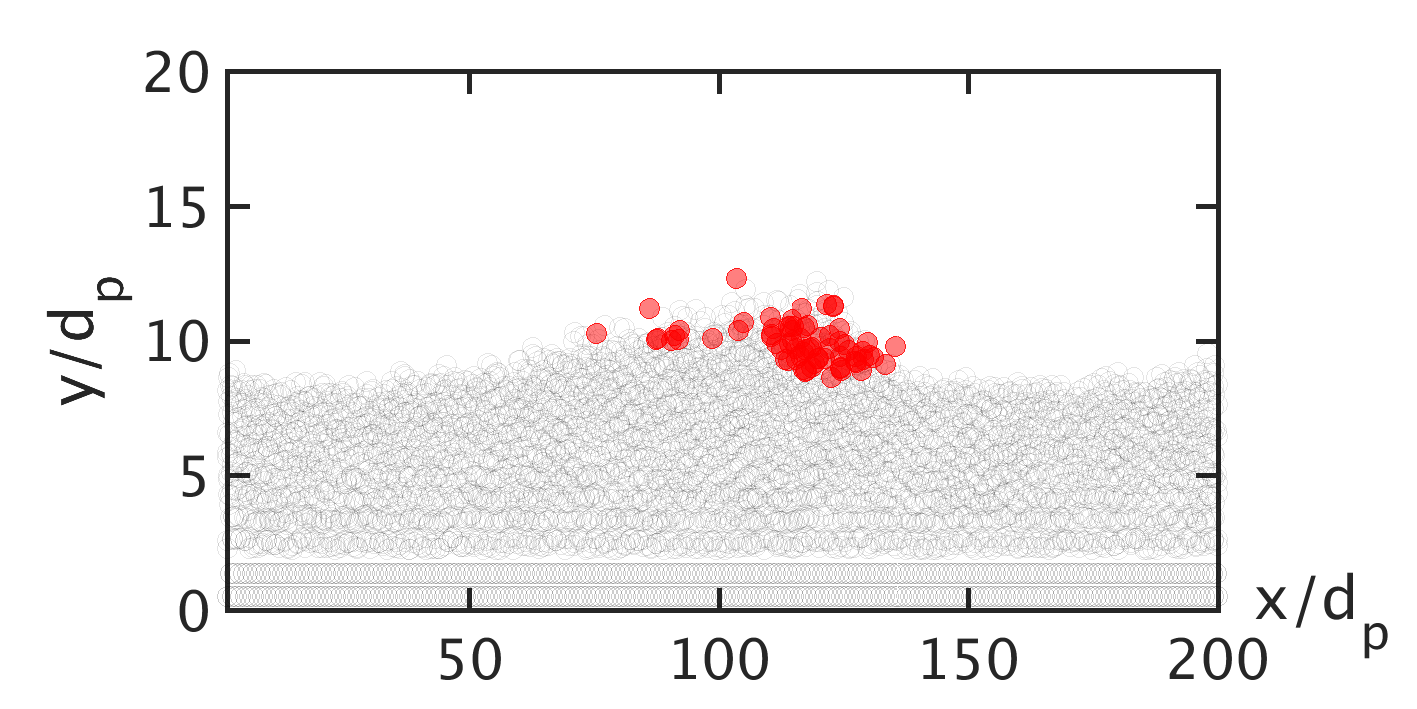}
  }
  \vspace{0.1 in}
  \subfloat[$t_0^* + 80$]{
  \includegraphics[width=0.45\textwidth]{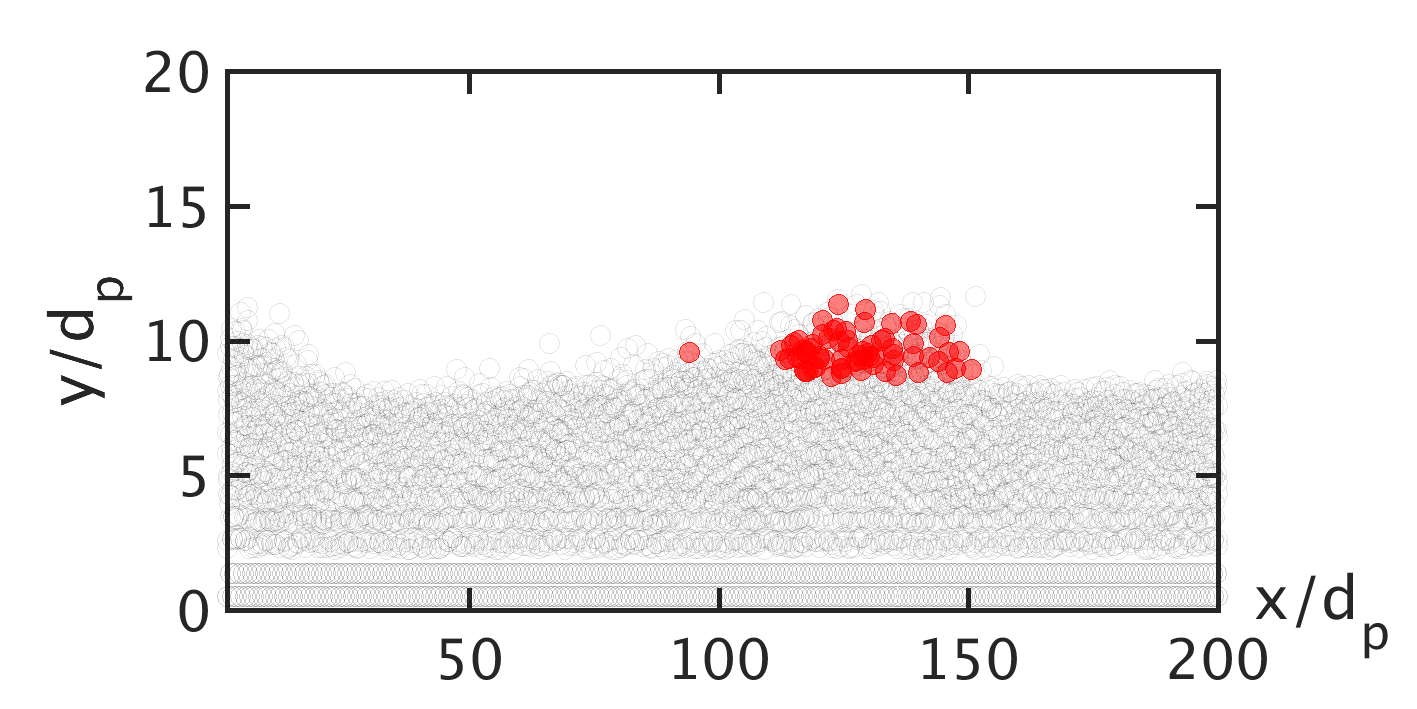}
  }
  \subfloat[$t_0^* + 120$]{
  \includegraphics[width=0.45\textwidth]{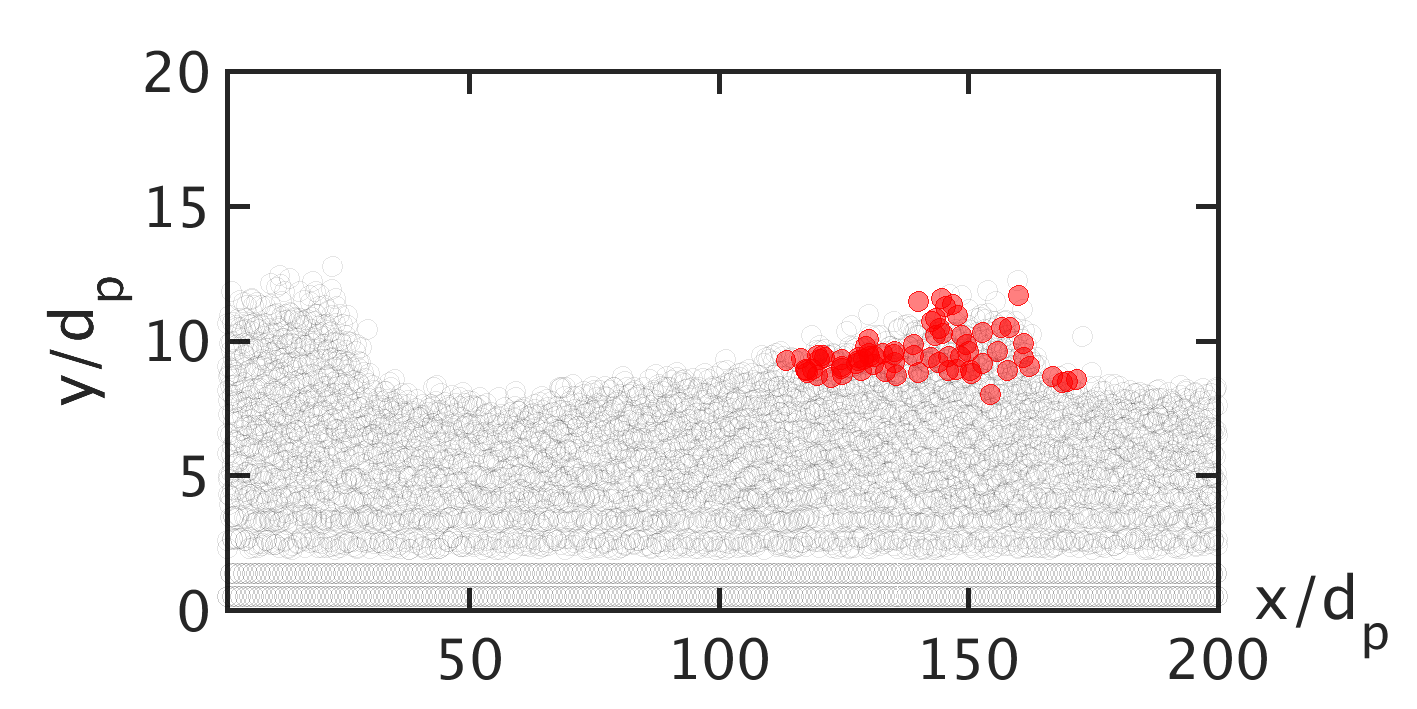}
  }
  \caption{The snapshots of the burial of the sediment particles on the crest of the bedform at bulk
    Reynolds number $\mathrm{Re_b}$~=~6,000. Each circle represents a sphere particle: red circles
    are highlighted particles; gray particles are non-highlighted. The first snapshot corresponds to
    non-dimensional time $t_0^*$~=~700.}
  \label{fig:bury-re-6000}
\end{figure}

The snapshots of the burial process of individual particles at bulk Reynolds number $\mathrm{Re_b}$
= 6,000 are demonstrated in Fig.~\ref{fig:bury-re-6000}. This study of the burial of sediment
particles aims to demonstrate the behavior of the deposited sediment particles during the bed
migration process after they jumped over the crest of the bedforms. This is complementary to the
previous study of the trajectories of the jumping particles. We selected 100 representative
particles on the surface at $t_0^* = 700$ and highlighted them in red (shade) color. It can be seen
in Fig.~\ref{fig:bury-re-6000}(a) that the highlighted particles are rolling and jumping on the bed
surface at the initial time step.  At $ t_0^* + 40$, these highlighted particles are deposited on
the same location ($x/d_p = 120$) at the lee side of the bedform, which is shown in
Fig.~\ref{fig:bury-re-6000}(b).  Comparing from the locations of the highlighted particles in
Fig.~\ref{fig:bury-re-6000}(b) and \ref{fig:bury-re-6000}(c), when the particles are buried in the
sediment bed, they do not move with the migration of the bedform. This is because the buried
particles are below the bed surface and the influence of fluid flow on them is negligible.  When the
bedform continues to move forward, shown in Fig.~\ref{fig:bury-re-6000}(d), some buried particles
are exposed on the bed surface because of flow erosion on the stoss side.  Once the previously
buried particles are exposed on the bed surface, they can be entrained in the fluid flow and move on
the surface of the bed. Compared with the observations in Fig.~\ref{fig:trace-re-6000}, the buried
particles do not move with the migration of the bedform, although the bedform progresses downstream
due to the entrainment and deposition of the particles on the bed surface. Visualization of the
burial process shows that only the particles on the bed surfaces are migrating along with the
bedform. This observation sheds light on the grain-level dynamics in bedform evolution and can
provide valuable insights for the numerical modeling. 

\section{Discussion on the averaged sediment transport quantities}
\label{sec:ap-averaged}

In addition to the characteristics of individual sediment particles, the averaged quantities of
sediment particles are detailed to investigate their spatial variation due to bedform
generation and migration. In this section, the sediment concentration, the sediment velocity, the
particle entrainment and deposition rates, and the bed and suspended loads are
investigated.  These quantities are averaged both in time and laterally in a moving frame of
reference to present the quantities as a function of the vertical direction, $y$, and the relative
longitudinal location $X=x-U_d t$ with respect to the bedform. 

\begin{figure}[htbp]
  \centering
  \subfloat[Case 1: $\mathrm{Re_b}$ = 6,000]{
  \includegraphics[width=0.45\textwidth]{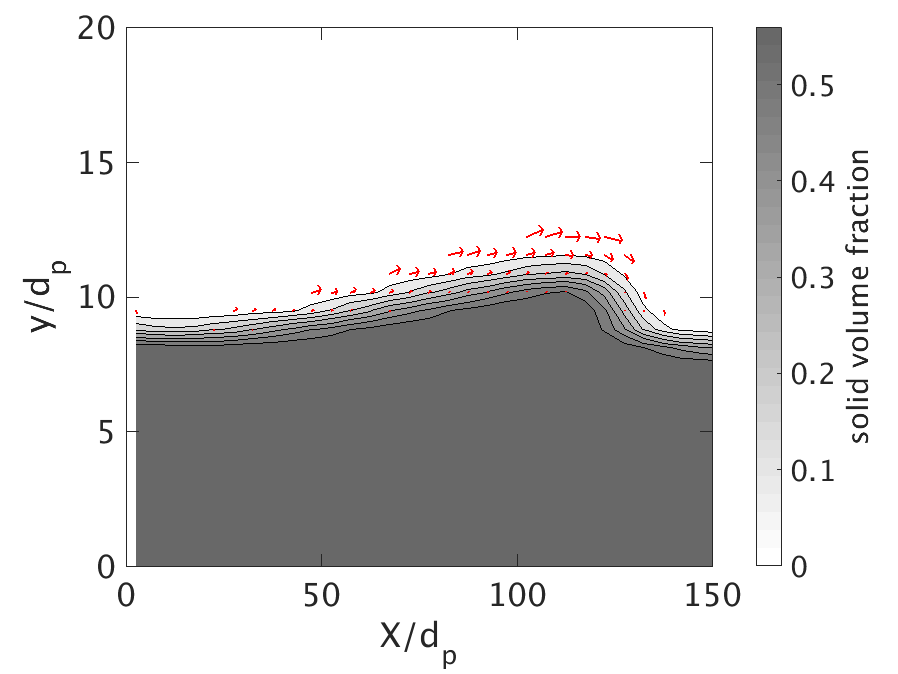}
  }
  \subfloat[Case 2: $\mathrm{Re_b}$ = 8,000]{
  \includegraphics[width=0.45\textwidth]{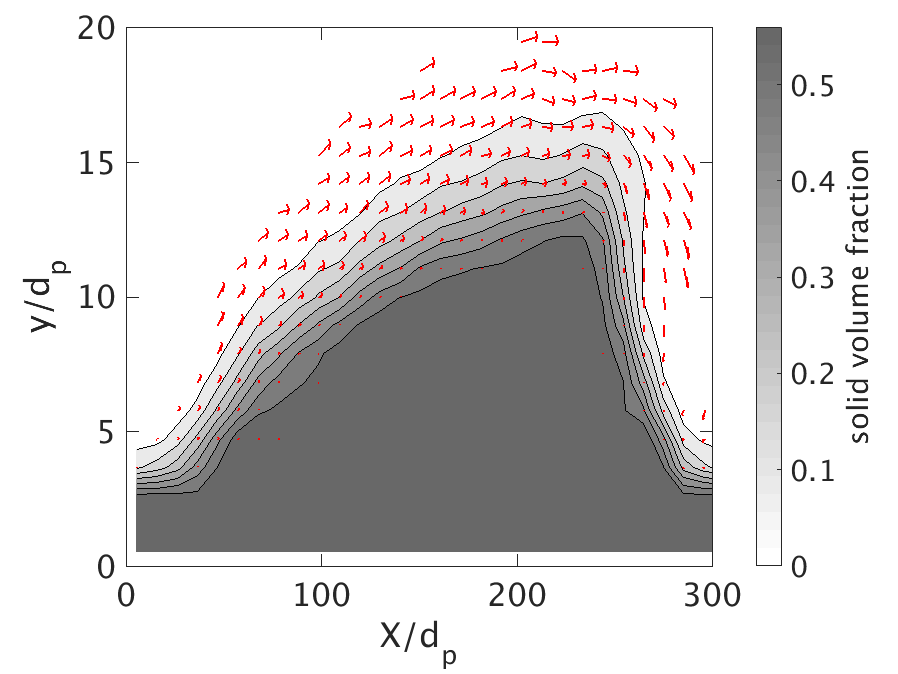}
  }
  \hspace{0.1in}
  \subfloat[Case 3: $\mathrm{Re_b}$ = 10,000]{
  \includegraphics[width=0.45\textwidth]{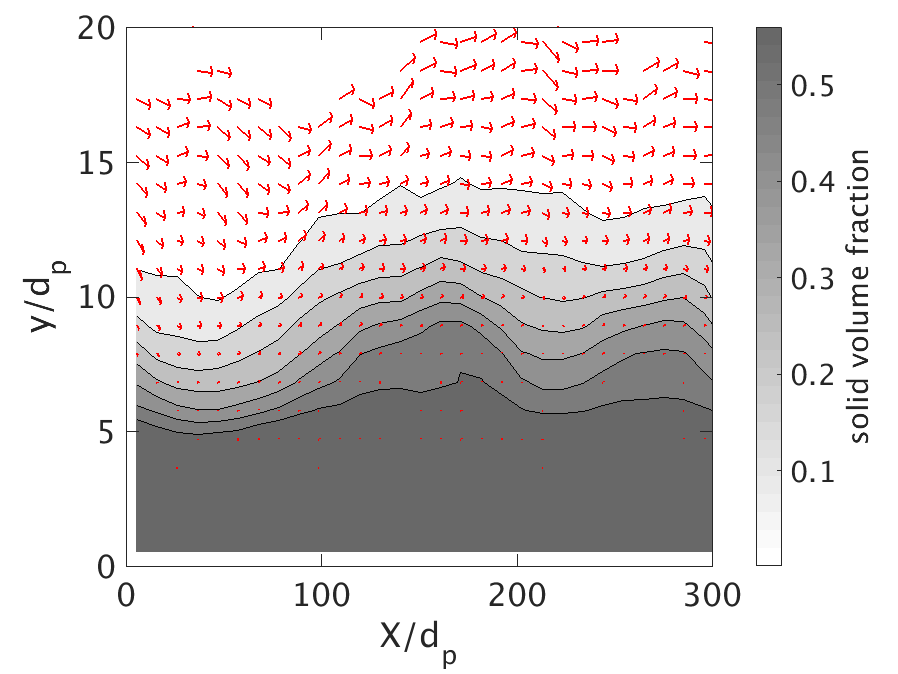}
  }
  \subfloat[Case 4: $\mathrm{Re_b}$ = 12,000]{
  \includegraphics[width=0.45\textwidth]{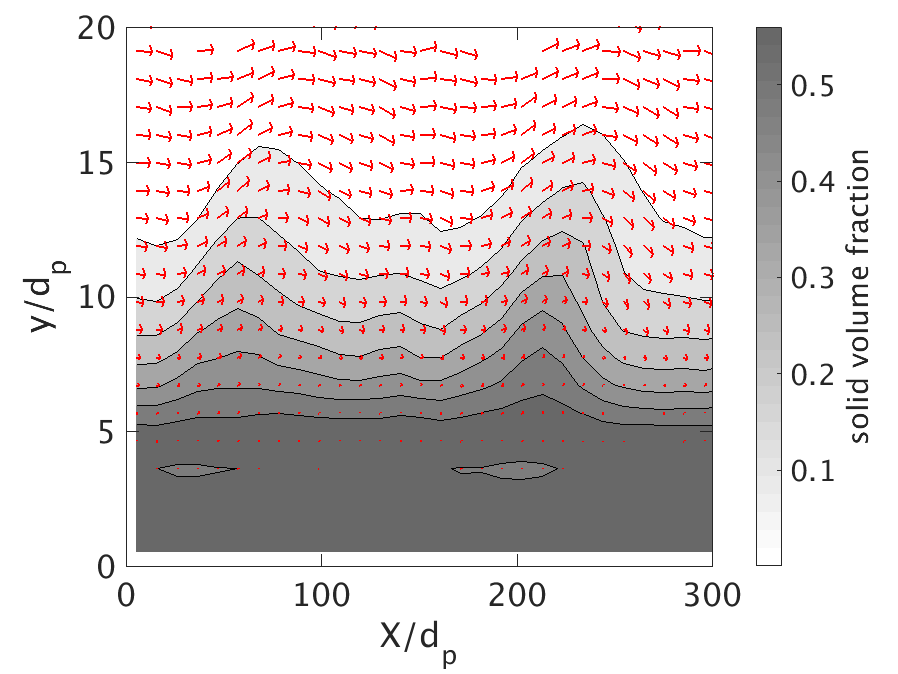}
  }
  \caption{Moving-frame averaged solid concentration contour at various Reynolds numbers. On top of
  the concentration contour, the arrows denote the moving-frame averaged velocity of sediment
  particle.}
  \label{fig:mfa-c}
\end{figure}

\begin{figure}[htbp] 
  \centering
  \subfloat[$\mathrm{Re_b} = 6,000$]{
    \includegraphics[width=0.45\textwidth]{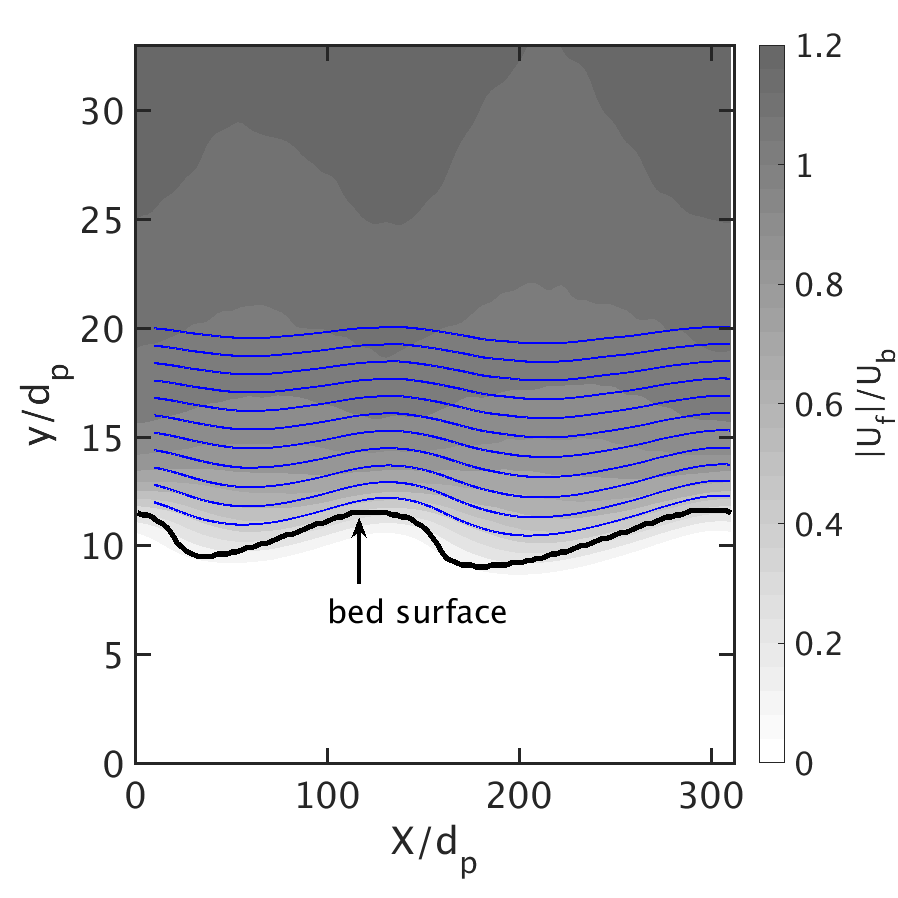}
  }
  \subfloat[$\mathrm{Re_b} = 8,000$]{
    \includegraphics[width=0.45\textwidth]{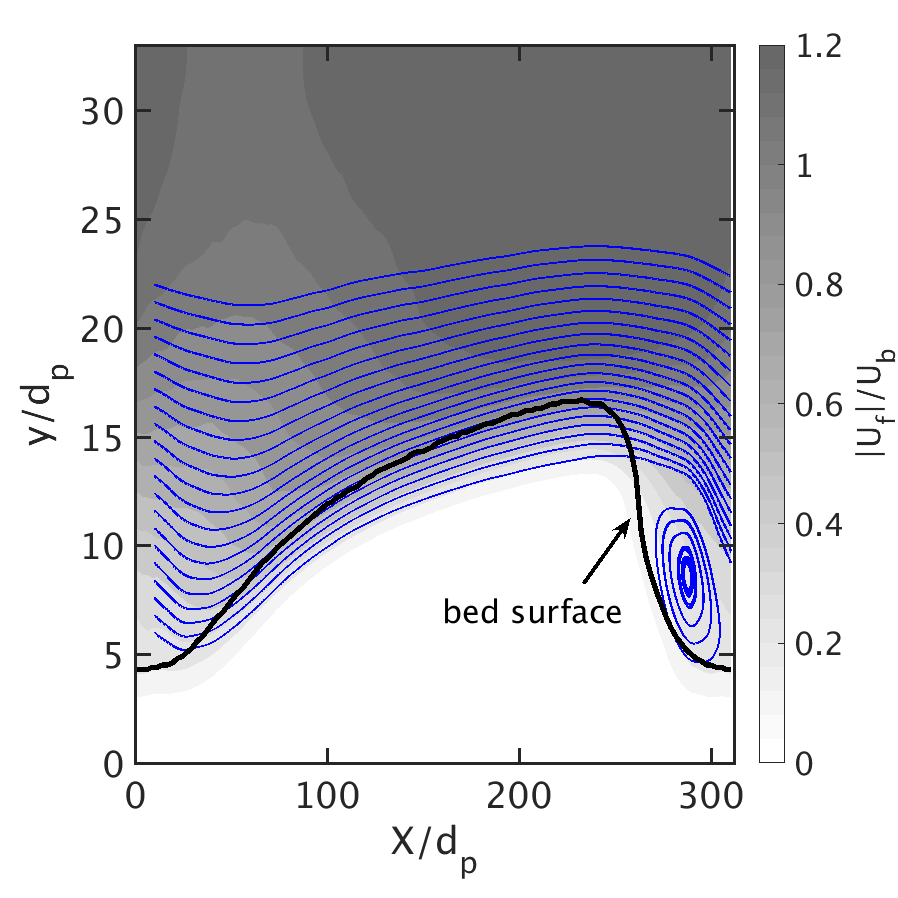}
  }
  \vspace{0.1 in}
  \subfloat[$\mathrm{Re_b} = 10,000$]{
    \includegraphics[width=0.45\textwidth]{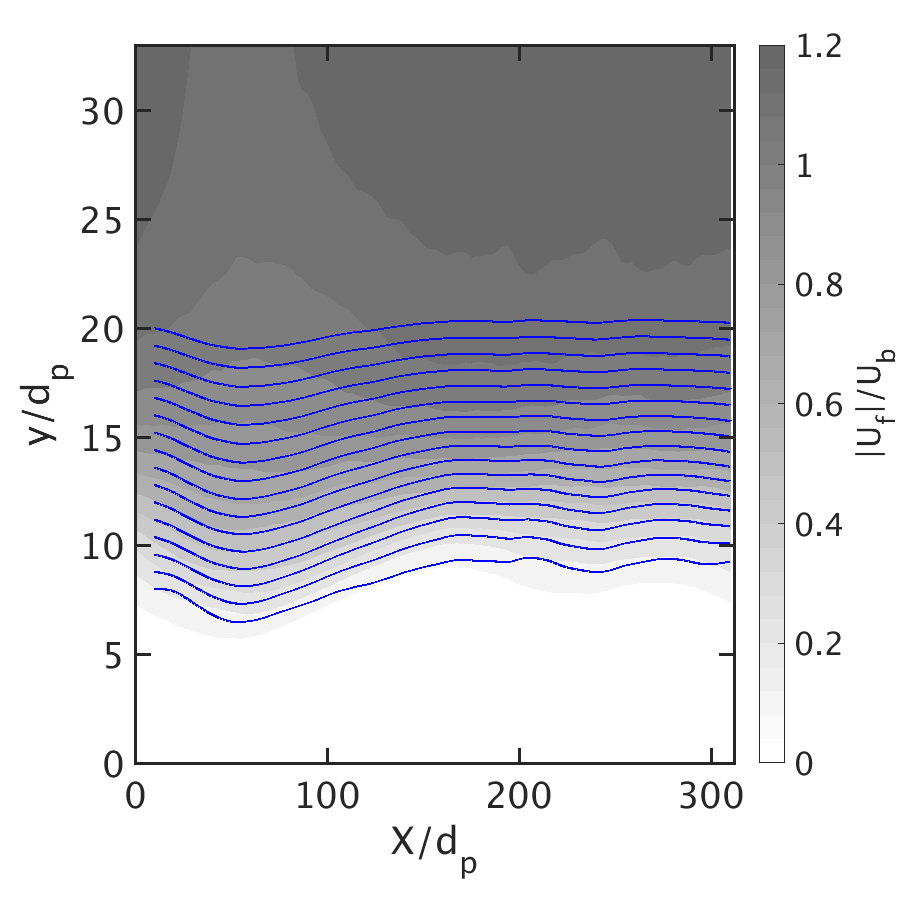}
  }
  \subfloat[$\mathrm{Re_b} = 12,000$]{
    \includegraphics[width=0.45\textwidth]{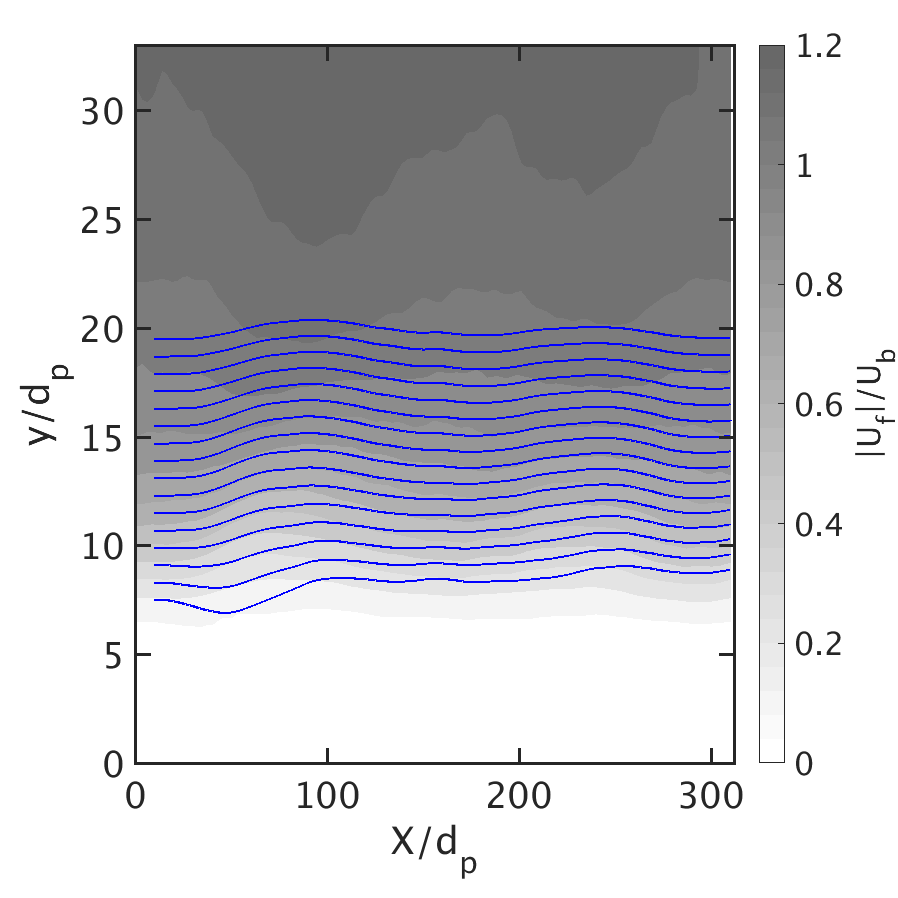}
  }
  \caption{ Moving-frame averaged flow velocity contour at various Reynolds numbers. On top of
  the flow velocity contour, the blue (solid) lines denote the streamline of moving-frame averaged
  velocity of fluid flow. The black (thick) lines denote the moving-frame averaged bed surface
  profiles.}
  \label{fig:flow-u-comp}
\end{figure}

The contours of the moving-averaged sediment concentration and velocity from $t^* \in [600,800]$ are
shown in Fig.~\ref{fig:mfa-c}. This demonstrates the variation of sediment concentration and
velocity at different bulk Reynolds numbers and sediment transport regimes. It can be seen in the
concentration contours that the sediment concentration is more diffusive at larger Reynolds numbers.
At $\mathrm{Re_b}$ = 6,000, the bedform migration velocity is stable, and the distance $\delta_s$
between the concentration isosurfaces of $\varepsilon_s =$ 0.1 and 0.5 is $2d_p$. At $\mathrm{Re_b}$
= 8,000, although the bedform migration velocity is stable, the saltation height of the sand particle
is increased, and thus the distance $\delta_s$ increases. When the bulk Reynolds number increases to
12,000, the suspended load becomes dominant, and there is a dramatic increase in the momentum
exchange between the fluid and sediment particle. Therefore, the distance between the concentration
isosurfaces $\delta_s$ can be as large as $10d_p$. It should be noted that the dune-like features in
Figs.~\ref{fig:mfa-c}(c) and~\ref{fig:mfa-c}(d) is because of the generation of unstable bedform.
Moreover, the magnitude and direction of particle velocity are plotted using arrows on top of the
concentration contour. This aims to demonstrate the averaged motion of sand particles and its
correlation to bed surface. It can be seen in both Fig.~\ref{fig:mfa-c}(a) and \ref{fig:mfa-c}(b)
that the particle velocities at the crest of both bedforms are larger than at other locations. It
can be also seen that the arrows of the particle velocity under the dune is too small and are not
visible in the figure. To help with the explanation of the particle downstream motion, we have also
plotted the moving-averaged fluid velocity in Fig.~\ref{fig:flow-u-comp}. The predictions of flow
velocity contour using LES--DEM are qualitatively consistent with the measurement in the
experimental study~\citep{NaqshbandEtal2014}, although the fluid flow data cannot be compared
directly with the experimental data. According to the flow velocity contour, the increase of
particle velocity is due to the increase of the fluid velocity at the bedform crest, which is also
consistent with the findings in previous experimental measurements~\citep{kostaschuk2004measuring,
KostaschukEtal2009}. When a larger bedform is generated at $\mathrm{Re_b} = 8,000$, the
recirculation of the fluid flow after the crest is captured in the simulations. Note that the
streamlines can be seen under the bed surface of $\varepsilon_s =$ 0.1 in
Fig.~\ref{fig:flow-u-comp}(b). This can be explained by the fact that the fluid flow moves with the
sediment particles in the bed load layer in the sediment bed.

\begin{figure}[htbp]
  \centering
  \includegraphics[width=0.45\textwidth]{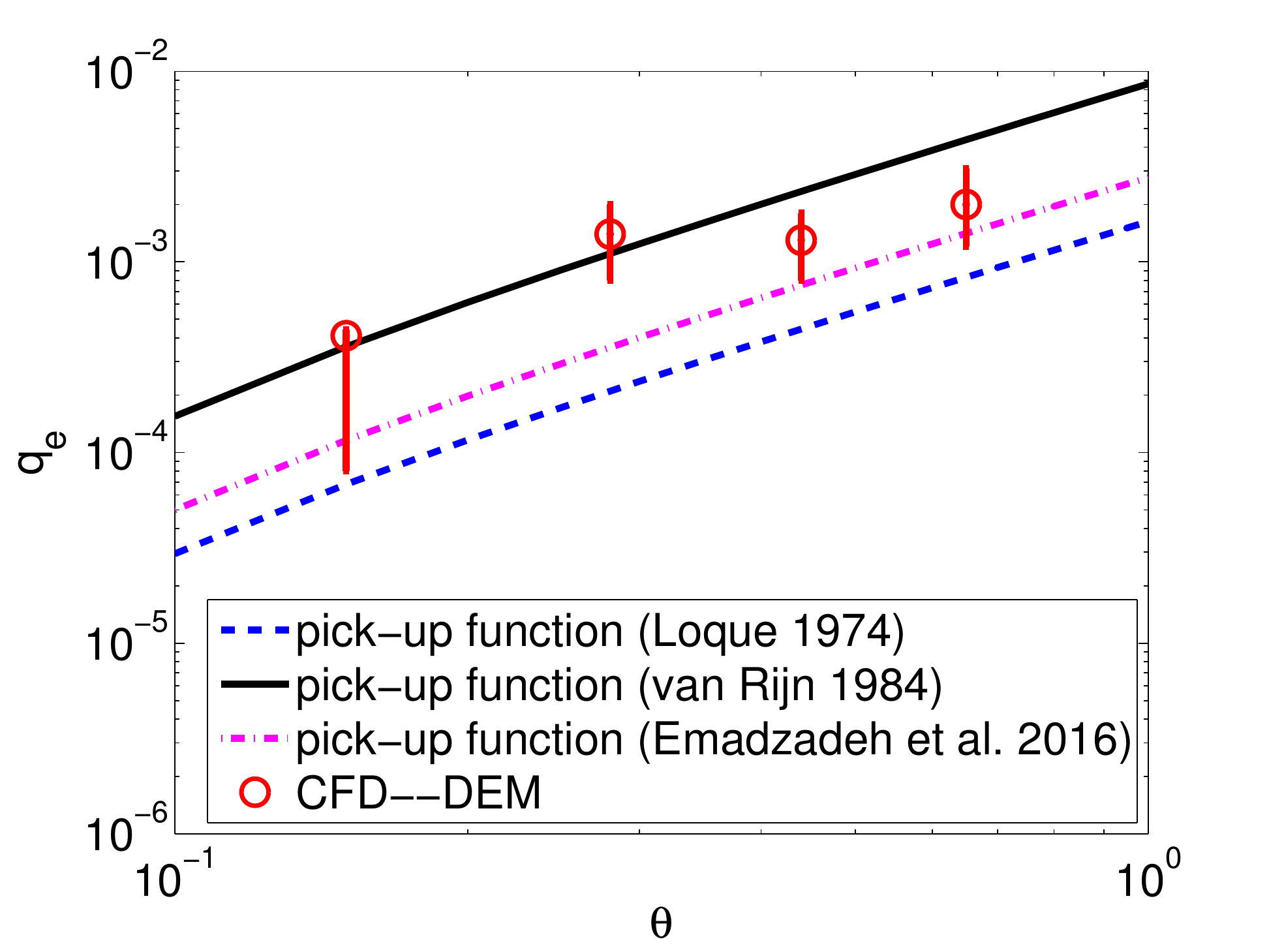}
  \caption{Horizontally averaged sediment entrainment rate during $t^*$~$\in$~$[700, 750]$ plotted as
    a function of Shields parameter. The markers indicate the entrainment rate obtained 
    on the isosurface of $\varepsilon_s$~=~0.1, and the error bars indicate the range of entrainment
    rates obtained at different isosurfaces ranging from $\varepsilon_s$~$\in$~$[0.05,0.5]$.}
  \label{fig:pick-up}
\end{figure}

In the vertical direction, the quantities of interest to study sediment transport are the sediment
entrainment and deposition, which are difficult to measure directly in experiments. Here we present
both entrainment rate $q_e$ and deposition rate $q_d$, which are a product of average sediment
concentration and particle velocity $\varepsilon_s u_{p,y}$, i.e., they are the volume flux of sand
particles per unit area.  Since the behavior of individual particles can be traced,
the entrainment rate is calculated using particles having positive vertical velocity; whereas the
deposition rate is obtained by using particles having negative vertical velocity. By using this
definition, the calculated entrainment rate is the sediment flux entrained in the flow; the
deposition rate is the sediment flux added to the bed. This is consistent with the definition of
sediment entrainment and deposition. The entrainment rate is plotted in Fig.~\ref{fig:pick-up}.  It
should be noted that the entrainment rate is not constant along the vertical or streamwise
directions. Therefore the sediment entrainment rates shown in the figure are a bed-averaged rate.
The markers in the figure are the entrainment rates at $\varepsilon_s = 0.1$, and the error bars
denote the range of the entrainment rates between the isosurfaces of sediment concentration
$\varepsilon_s=0.05$ to 0.5. Empirical functions for entrainment based on flume data are also shown
in the figure \citep{luque1974erosion, rijn1984spu, emadzadeh16ms}. In general, the bed-averaged
LES--DEM derived values fall within the range of these empirical formulas and are closest in value
to that of \cite{rijn1984spu}.

\begin{figure}[htbp]
  \centering
  \subfloat[Case 1: $\mathrm{Re_b}$ = 6,000]{
    \includegraphics[width=0.45\textwidth]{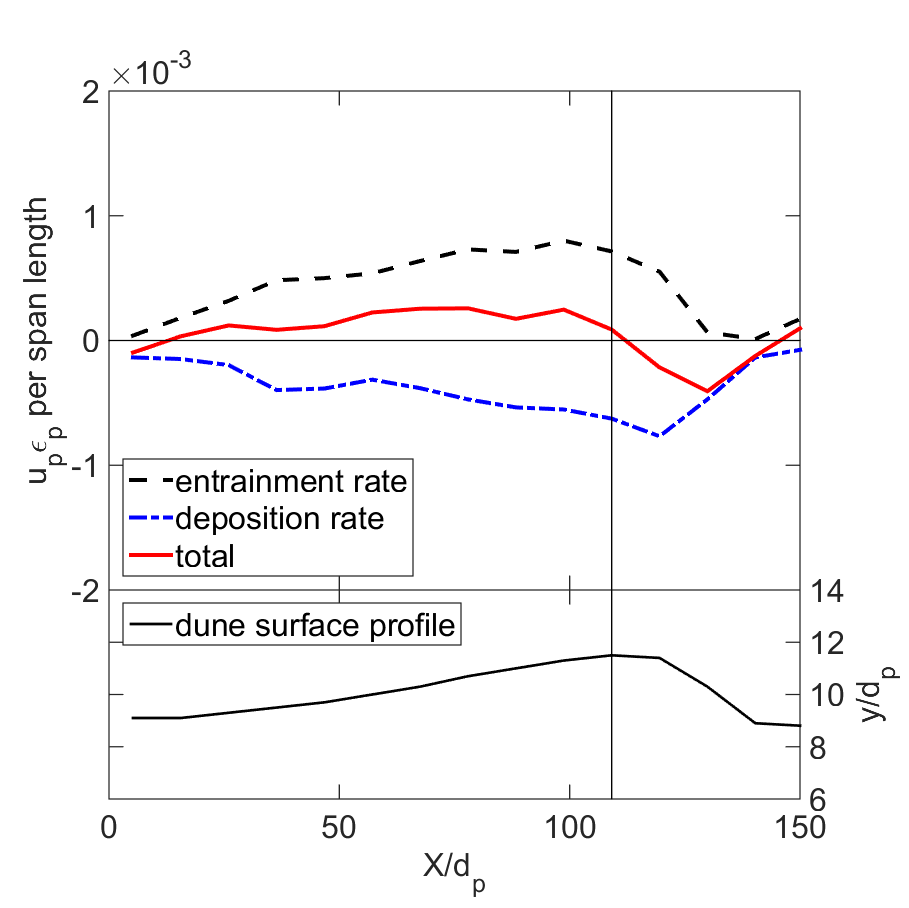}
  }
  \subfloat[Case 2: $\mathrm{Re_b}$ = 8,000]{
    \includegraphics[width=0.45\textwidth]{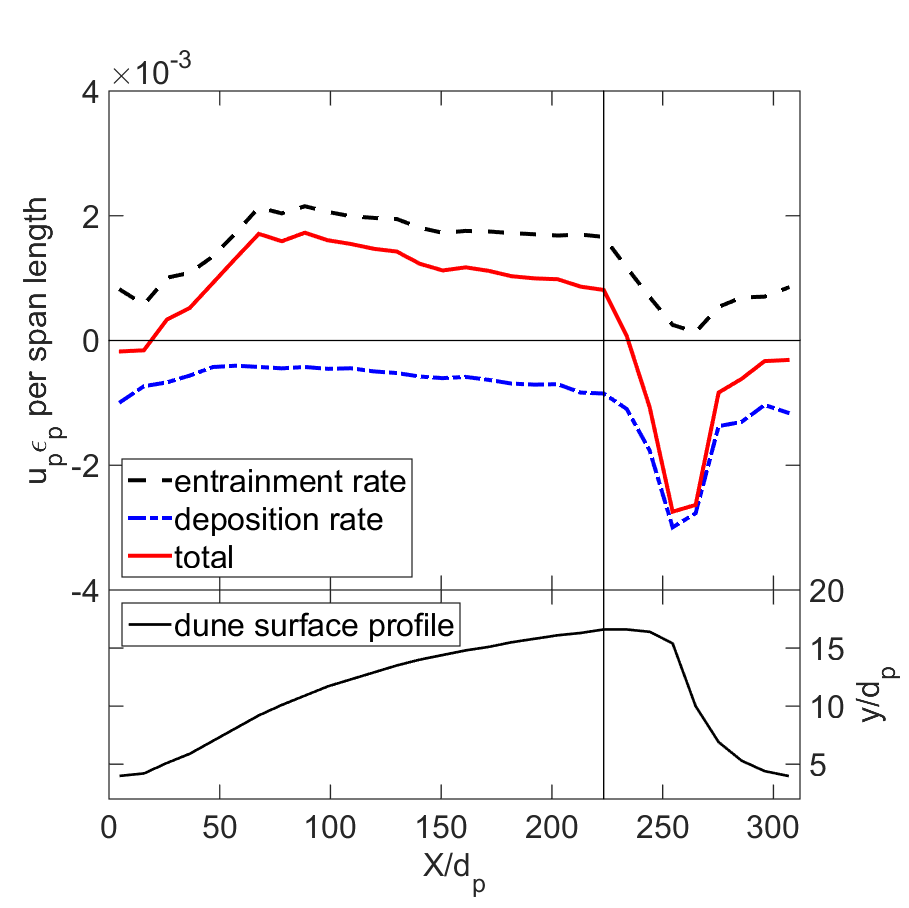}
  }
  \caption{Comparison between the entrainment and deposition of sediment particles at
    $\mathrm{Re_b}$~=~6,000 and 8,000. The lateral-averaged bed profiles are plotted below to
    demonstrate the location of the stoss side and lee side. The vertical solid line in the figure
    indicates the location of the peak of the bedform. The range of the $x$-axis in each panel is
    consistent with the bedform wavelength.}
  \label{fig:mfa-ed-load}
\end{figure}

To demonstrate the variation of the vertical sediment transport, the entrainment rate and deposition
rate at the bed surface, $\varepsilon_{s} = 0.1$, are plotted as a function of the streamwise
coordinate in Fig.~\ref{fig:mfa-ed-load}. When the bedform is stable in the moving frame of reference
($\mathrm{Re_b}$ = 6,000 and 8,000), the total sediment flux in the vertical direction is above zero
on the stoss side of the bedform, and becomes negative on the lee side. That is not to say that only
entrainment occurs on the stoss and deposition in the wake, but rather that there is net entrainment
over the stoss and net deposition in the wake. Both entrainment and deposition rate increase with
Reynolds number.

In general, the total streamwise sediment transport rate, $q_{t}$ or
$q_{t}^{*}=q_{t}/d_{p}\sqrt{gR_{s}d_{p}}$ [-], can be found by integrating the particle velocities
and concentrations over the total thickness of the domain and then averaging this over the length of
the domain. These values are given in Table \ref{tab:dune-transport}. However, it is also of
interest to decompose this total downstream transport rate into bed, $q_{b}$, and suspended,
$q_{s}$, load fractions. To this end, particles moving above the isosurface of the critical sediment
volume fraction $\varepsilon_{s,cr}$ are considered as suspended load and those moving below this
isosurface are considered as bed load. The critical sediment volume fraction to determine the
suspended load is proposed in~\cite{rijn84se2}:
  \begin{equation}
    \varepsilon_{s,cr} = 0.03\frac{d_p}{H_d}\frac{T^{1.5}}{D*^{0.3}},
    \label{eqn:epsilon-cr}
  \end{equation}
  where non-dimensional particle diameters $D_*$ is: 
  \begin{equation}
    D_* = d_p \left[ \frac{(s-1)g}{\nu^2} \right]^{1/3},
    \label{eqn:d_star}
  \end{equation}
  and the transport stage parameter $T$ is:
  \begin{equation}
    T = \frac{(u_{*}')^2-(u_{*,cr})^2}{(u_{*,cr})^2}. 
    \label{eqn:teqn}
  \end{equation}
Based on these criteria, the simulations show an increase in the overall suspended load fraction,
$\alpha_{s}=q_{s}/q_{t}$, from 0.14 in Case 1, up to $\alpha_{s} = 0.66$ at Case 4. Values for
$\alpha_{s}$ are given in Table \ref{tab:dune-transport}. The overall suspended load fraction
obtained in the present simulations are consistent with the summary of experimental results
in~\citep{rijn84se2}. Note that the total domain-averaged transport rates were used in the
calculation of $\alpha_{s}$.

\begin{figure}[htbp]
  \centering
  \subfloat[Case 1: $\mathrm{Re_b}$ = 6,000]{
    \includegraphics[width=0.45\textwidth]{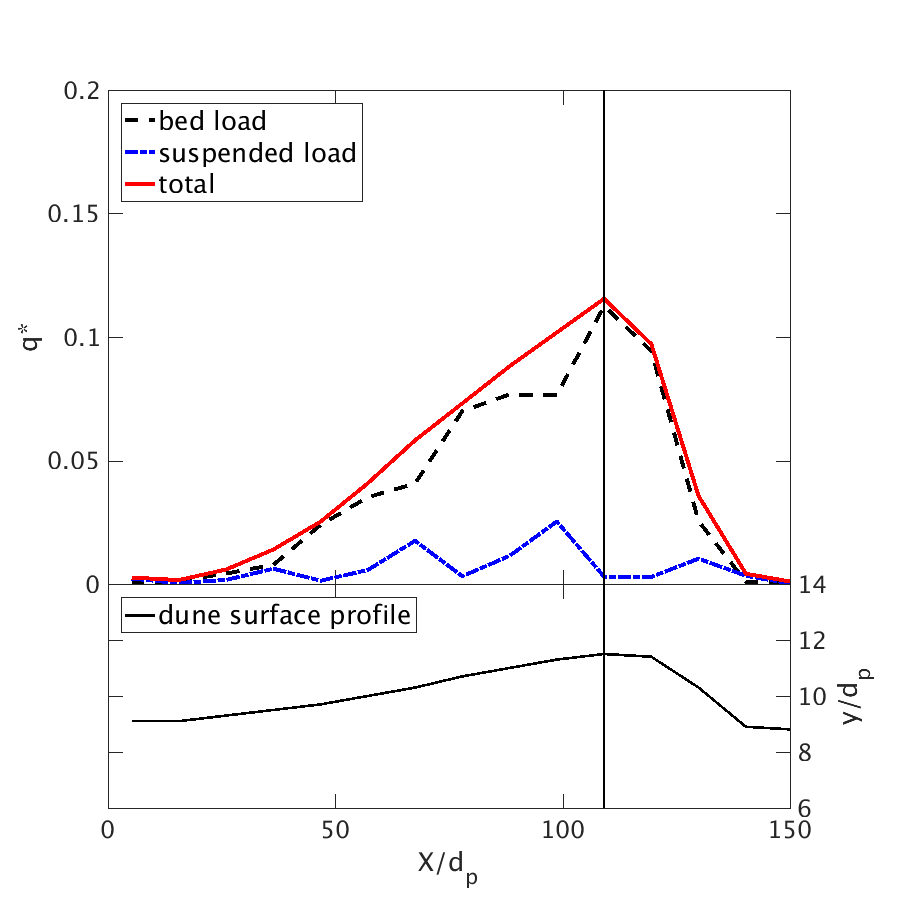}
  }
  \subfloat[Case 2: $\mathrm{Re_b}$ = 8,000]{
    \includegraphics[width=0.45\textwidth]{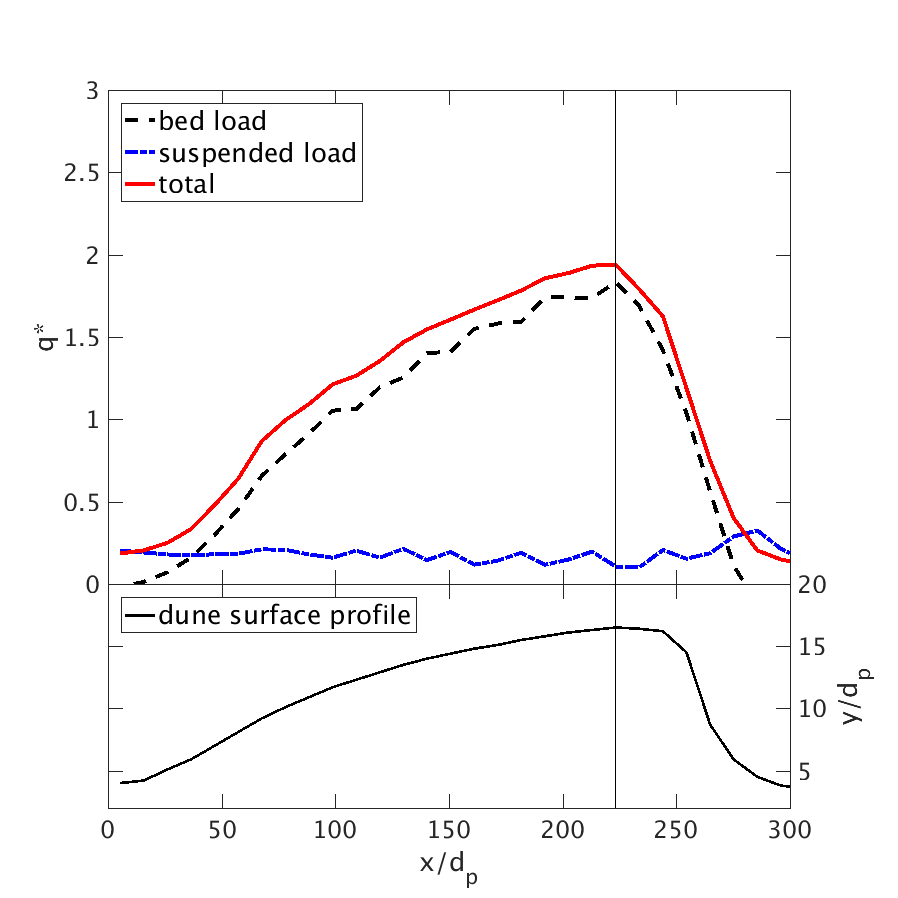}
  }
  \caption{Comparison between the bed load and suspended load of sediment particles at bulk Reynolds
    number $\mathrm{Re_b}$~=~6,000 and 8,000 during $t^* \in [600,800]$. The lateral-averaged bed
    profiles are plotted below to demonstrate the location of the stoss side and lee side.  The
    vertical line in the figure indicates the location of the peak of the bedform.  The range of the
    $x$-axis in each panel is consistent with the bedform wavelength.}
  \label{fig:mfa-bs-load}
\end{figure}

We can also look at the distribution of the non-dimensional total, bed, and suspended loads over the
length of a bedform in a moving frame of reference. Such distributions are shown in
Fig.~\ref{fig:mfa-bs-load}. The results are obtained by using the moving-averaged data
from $t^* \in [600,800]$ for Case 1 and 2. The moving-averaged bed profiles is also plotted in
Fig.~\ref{fig:mfa-bs-load} to demonstrate the correlation between of transport rates and the bed
profile. The most striking feature of the load distribution is that $q_{t}^{*}$ follows the general
bedform shape. Total transport rates increase moving from the preceding trough up and over the stoss
of the bedform. Peak transport is reached near the crest of the bedform and then rapidly falls off
in the wake. At $\mathrm{Re_b}$ = 6,000, bed load is by far the dominant contributor to the total
transport rate. Additionally, no transport occurs in the trough between sequent bedforms. That is, all
sediment being transported in the domain is tied up in the movement of the bedforms. However, at
$\mathrm{Re_b}$ = 8,000, a far greater percentage of the sediment being moved travels in suspension.
Interestingly, the bed load transport rate at $\mathrm{Re_b}$ = 8,000 is fairly constant in the
streamwise direction over the length of the stoss.  The peaking nature of the total transport rate
over the length of the bedform, therefore, comes from the increasing contribution of suspended load
over the stoss as the particle velocities increase. As with the $\mathrm{Re_b}$ = 6,000 case, bed
load transport in the trough between sequent bedforms goes to zero.  However, the total transport
rate remains positive in the trough due to the bypass contributions from suspended load
(Fig.~\ref{fig:mfa-bs-load}).

One additional way to estimate bed load transport rates is through use of the bedform height,
celerity, and porosity \citep[e.g.,][]{SimonsEtal1965, McElroyMohrig2009}. We calculated the bedform
transport rate, $q_{\mathit{bf}}$, as:
\begin{equation}
q_{\mathit{bf}}=(1-\phi)U_{d}\frac{H_{d}}{2}
\label{eq:bedformq}
\end{equation}
where $\phi$ is the average porosity of the bed (taken to be $\phi=0.5$); note
Eq.~(\ref{eq:bedformq}) assumes the beforms are triangular in shape. By using Eq.~(10), we compared
the calculation of the bedform transport rate with our previous definition in which the particles
moving below the isosurface of the critical concentration $\varepsilon_{s,cr}$ are bed load at
$\mathrm{Re_b}$= 6,000 and 8,000. Non-dimensional values of $q_{\mathit{bf}}$ are given in Table
\ref{tab:dune-transport} for Cases~1 and~2 using values of $U_{d}$ and $H_{d}$ also reported in the
same table. Eq.~(\ref{eq:bedformq}) did well in estimating the total transport load in Case 1
($\mathrm{Re_b}$= 6,000); $q_{\mathit{bf}}^{*}=0.06$ compared to $q_{t}^{*}=0.05$. This result is
reasonable seeing that the total transport rates go to zero in the troughs for Case 1
(Fig.~\ref{fig:mfa-bs-load}(a)). The value of $q_{\mathit{bf}}^{*}=0.93$ in Case 2 is larger than
that in Case 1, but is smaller than the total load determined from the particle motions,
$q_{t}^{*}=1.12$. The reduction in $q_{\mathit{bf}}$ relative to $q_{t}^{*}$ is likely an outcome of
the fact that $q_{t}^{*}$ does not go to zero in the trough for the $\mathrm{Re_b}$= 8,000 run
(Fig.~\ref{fig:mfa-bs-load}(b)). Interestingly, while $q_{\mathit{bf}}^{*}$ was smaller than the
total load at this condition, $q_{\mathit{bf}}^{*}$ was higher than the bed load transport rate,
$q_{b}^{*}=0.52$. This implies that some of the material counted as suspended load, in our
definition of suspended load, was contributing to the migration of the bedform.  The suspended load
can be determined from a difference between the total load and bedform transport rate, and the
difference $q_{s}^{*}=q_{t}^{*}-q_{\mathit{bf}}^{*}$, yields a suspended load fraction at
$\mathrm{Re_b}$= 8,000 of $\alpha_{s}=0.17$. This is consistent with the of $\alpha_{s}$ value
determined using our definition of suspended load as particles moving above the isosurface of
$\varepsilon_{s,cr}$.

\begin{figure}[htbp]
  \centering
  \subfloat[Case 1: $\mathrm{Re_b}$ = 6,000]{
  \includegraphics[width=0.45\textwidth]{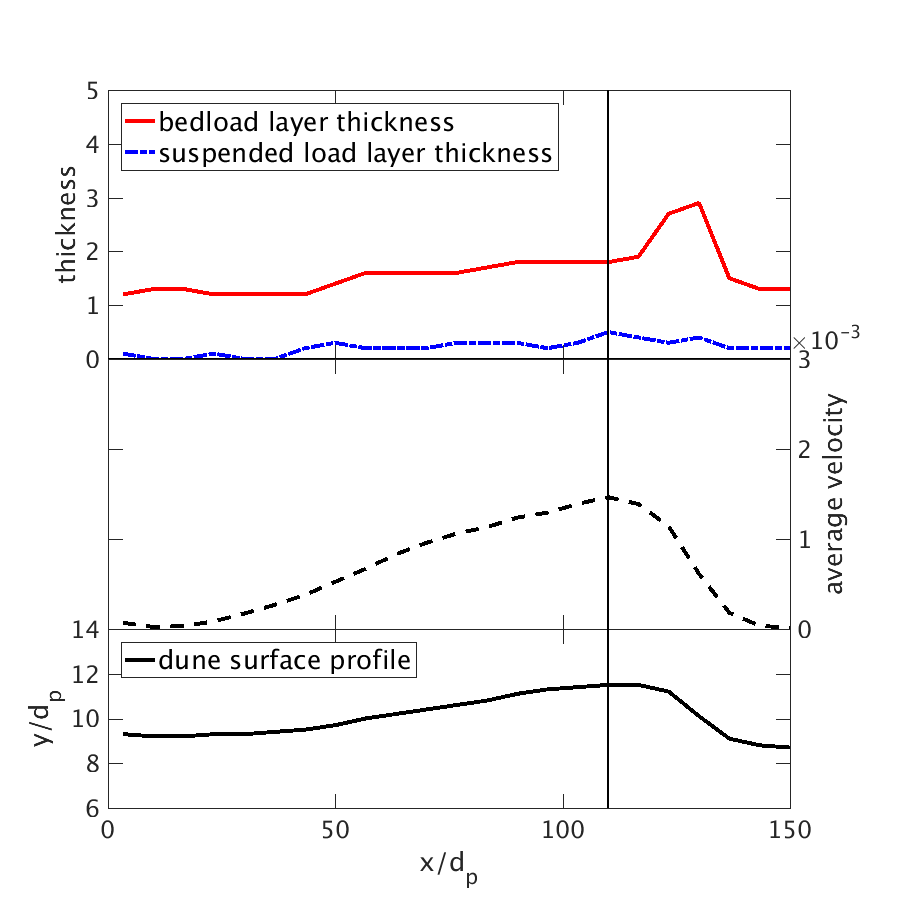}
  }
  \subfloat[Case 2: $\mathrm{Re_b}$ = 8,000]{
  \includegraphics[width=0.45\textwidth]{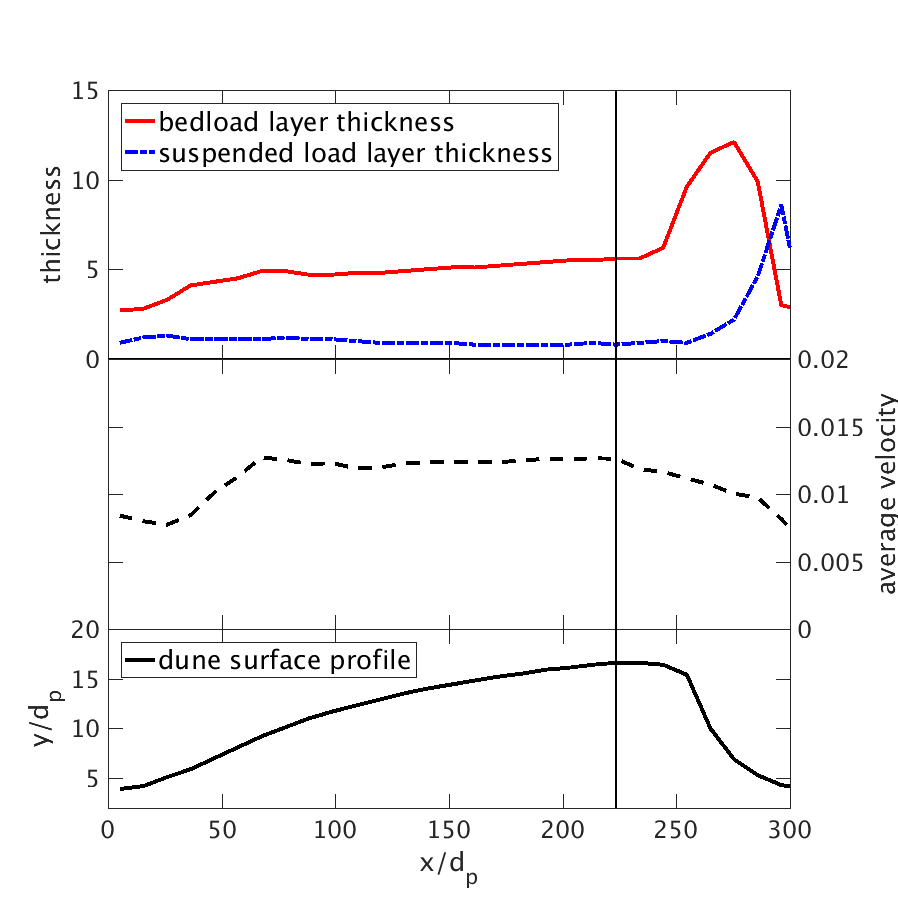}
  }
  \caption{Comparison between the thickness of bed load/suspended load layer and sediment particle
    velocity at $\mathrm{Re_b}$~=~6,000 and 8,000 during $t^* \in [600,800]$. The
    range of the $x$-axis in each panel is consistent with the bedform wavelength.}
  \label{fig:mfa-thickness}
\end{figure}

The thickness of both the bed and suspended load layers are shown as a function of the streamwise
coordinate in Fig.~\ref{fig:mfa-thickness}. The height of the bed load layer is defined as the
distance of the isosurfaces of sediment concentrations $\varepsilon_s$ between the critical value
$\varepsilon_{s,cr}$ and 0.5, and the height of suspended load layer is the distance of the
concentration isosurfaces between 0.01 and the critical value $\varepsilon_{s,cr}$.  The definition
to determine the bed load layer thickness is consistent with our previous definition in the
calculation of bed load flux. The bed surface profiles are also plotted to help illustrate the
location of peaks of the thickness. It can be seen in Fig.~\ref{fig:mfa-thickness} that the maximum
heights of both bed and suspended load layers are not at the peak of the bedform but on the lee side
of the bedform. The increase of the bed load layer thickness is because the slope of the bedform on
the lee side is significantly larger than the stoss side, and the thickness in the vertical
direction increases due to the increase of the slope; whereas the increase of suspended load layer
thickness at $\mathrm{Re_b}$ = 8,000 is due to particle suspension in the recirculation region after
peak of bedform. It can be also seen in the figure that both the thickness of the bed load layer and
suspended load layer is increasing when the bulk Reynolds number increases. The moving-frame
averaged particle velocity is also plotted as a function of streamwise coordinate in
Fig.~\ref{fig:mfa-thickness}. It can be seen in the figure that the trend of average particle
velocity is consistent with the bed profiles, which is consistent with the trend of sediment
transport rate in Fig.~\ref{fig:mfa-bs-load}.

\section{Conclusion}

In this study, we have investigated the individual particle motions associated with self-generated
bedforms at different bulk Reynolds numbers, or transport stages, using \emph{SediFoam}. The
simulations presented cover a range of transport conditions from bedform inception at transport
stage of $\theta/\theta_{cr}=4.6$ to washout conditions (at approximately $\theta/\theta_{cr}>10$
and $w_{s}/u_{*}' \approx 1$). Calculation of the individual particle statistics showed that washout
starts to occur when the jump length of the individual particle starts to exceed the wavelength of
the bedforms and computational domain. This also coincided with a transition from bed load dominated
transport to suspended load dominated transport.  The current LES--DEM model can complement the
experimental results by providing much detailed information of the system, e.g., individual particle
trajectories, statistics of particle migration behaviour, detailed flow field, and fluid--particle
interactions, among others. With the LES--DEM simulation results and the insights generated therein,
the closure terms in the two-fluid models or hydro-morphodynamic models can be improved, which can
contribute to the numerical modeling in the sediment transport problems in engineering scales.

When bedforms were present, both bed and suspended load increased over the stoss side of the
bedforms; reaching a peak in the individual particle and total transport just at or before the crest
of the bedforms. For the lower transport conditions ($\theta/\theta_{cr}=4.6$ and $w_{s}/u_{*}'=1.88$),
material moved almost exclusively in bed load, and nearly 100\% of the material transported from and
over the stoss was trapped in the lee-side wake. No material bypassed the bedform. At this condition,
the total sediment transport rate was well approximated by the bedform transport rate calculated
using the bedform height, celerity, and bed porosity (Eq.~(\ref{eq:bedformq})). As contributions
from suspended load increased ($\theta/\theta_{cr}=8.5$ and $w_{s}/u_{*}'=1.33$), a fraction of the
sediment was found to bypass the trough of the bedform. This resulted in a positive sediment
transport rate through the trough. One result of this was that the transport rate predicted from the
bedform properties underestimated the total transport rate. In this case, the difference between the
total load and sediment transport rate associated with bedform migration was not equal to the
suspended load transport rate. In our definition of suspended load, material can be picked up in
suspension and deposited back down onto the sediment bed in a way that contributes to the overall
size and migration speed of the bedform. Hence, calculating the transport rate from the bedform
properties included contributions from both bed and suspended load. 

In this study we do not examine the fluid mechanics of the system or the links between coherent flow
and the sediment motion. Instead, we have highlighted the ability of LES--DEM methods to obtain the
detailed particle motion data that cannot, at least as of yet, be obtained through field or
laboratory studies of sand beds. The type and amount of data that can be generated with such methods
is rather staggering, and the presence of mobile bedforms presents added challenges with regard to
presenting the detailed data in a meaningful way. In this paper, we have chosen to present data on
individual particle jump lengths, particle entrainment into suspension, and the fractionation
between bed and suspended load within a moving frame of reference that is tied to the celerity of
the bedforms. We demonstrate that the LES--DEM method can be used to both investigate individual
particle dynamics and extract key sediment transport quantities needed to bolster modeling of mobile
beds in coarser resolution simulations. For example, use of the LES--DEM method allowed for the
evaluation of the particle pickup rates over the heterogeneous stress field of the mobile bed
surfaces. Depending on how the bed load and suspended load layers are conceptualized as existing,
one could use this type of information to further develop models of suspended load entrainment under
a range of condition.

While methods such as \emph{SediFoam} are opening up new doors for the exploration of detailed
particle dynamics in sediment transport research, such simulations are still limited by
computational resources. For example, we were not able to generate stable bedforms at
$\mathrm{Re_b>8,000}$ (or $\theta/\theta_{cr}>10$). At this point it is difficult to say whether the
bedforms were washed out because the mode of transport switched unequivocally to suspended load or
because the domain became smaller than the jump length of the particles. For example, at
$\mathrm{Re_b=}$10,000 and 12,000, correlated waves of suspended load could be observed moving
through the domain (Figs.~\ref{fig:mfa-c}(c) and~\ref{fig:mfa-c}(d)). Perhaps such waves would lead
to longer wavelength bedforms given enough space. However, doubling the size of the domain would
further increase the number of particles and the mesh size required to resolve the flow field,
leading to increased computational costs. Another avenue left to push towards in the application of
LES--DEM methods for sand bed research is the incorporation of deeper flows, the presence of a free
surface, and the use of a grain size distribution within the bed. The form lift for
realistic particles is also important and contributes to the uncertaities of the results. As such
advances come to fruition, it is possible to conceive of parametric numerical simulations being run
to better refine our understanding of the particle physics involved in the movement of sand through
rivers.

\section*{Acknowledgments}
The computational resources used for this project are provided by the Advanced Research Computing
(ARC) of Virginia Tech, which is gratefully acknowledged.  RS gratefully acknowledges partial
funding of graduate research assistantship from the Institute for Critical Technology and Applied
Science (ICTAS, grant number 175258) in this effort. We thank the anonymous reviewers for their
comments, which significantly improve the quality of the paper. To access data used in this paper,
please contact the corresponding author.

\section*{Reference}
\bibliographystyle{elsarticle-harv}
\bibliography{references}

\appendix
\setcounter{secnumdepth}{0}
\renewcommand\thefigure{A.\arabic{figure}}
\renewcommand\thetable{A.\arabic{table}}
\setcounter{figure}{0}
\setcounter{table}{0}
\section*{Appendix: Grain Avalanching Test}

If the angle of the lee side of the bedform is larger than the angle of repose, the sediment particles
are falling along the lee side to the bottom. In the present simulations, the motions of individual
sediment particles are captured, and there is no ad-hoc modeling of the grain avalanching. In this
section, the grain avalanching test is performed to validate that \emph{SediFoam} can capture the
repose angle of the sediment particles.  

The parameters of the computational domain and mesh resolution of the grain avalanching test are
shown in Table~\ref{tab:param-avalanching}. The properties of fluid and sediment particles are
consistent with those detailed in Table~\ref{tab:param-all}. The fluid flow is quiescent
initially. A separate simulation of particle settling is performed to obtain an initial
configuration of the particles, and the particles fall from random positions under gravity with
inter-particle collisions. To provide a bottom boundary condition for the moving particles, the
bottom wall is covered with one layer of fixed particles with random perturbations.
The physical parameters of the simulation are also shown in Table~\ref{tab:param-all}.
The particle diameter is 0.5~mm and initial hight and width of the particle pile is 15~mm. The
velocity of the sediment particles is smaller than $1\times10^{-6}$~m/s after 0.6~s. The time step
is $5\times10^{-4}$~s for the fluid flow and $1\times10^{-6}$~s for the particles.

\begin{figure}[htbp]
  \centering
  \subfloat[T = 0.02s]{
    \includegraphics[width=0.45\textwidth]{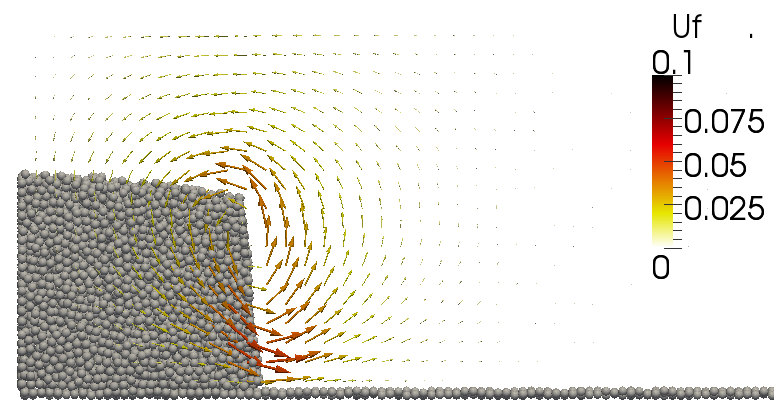}
  }
  \subfloat[T = 0.10s]{
    \includegraphics[width=0.45\textwidth]{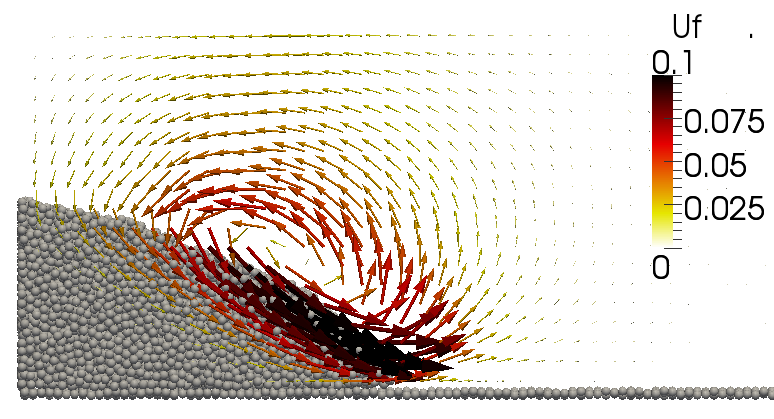}
  }
  \vspace{0.02in}
  \subfloat[T = 0.20s]{
    \includegraphics[width=0.45\textwidth]{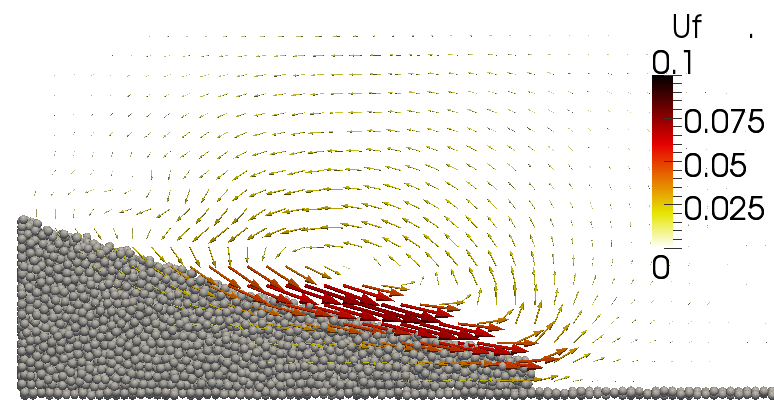}
  }
  \subfloat[T = 0.60s]{
    \includegraphics[width=0.45\textwidth]{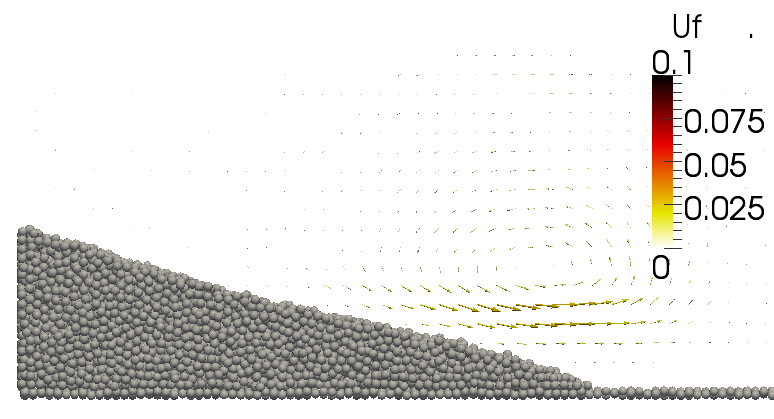}
  }
  \caption{The snapshots of sediment particles arrangement obtained in the avalanching test at
  different times. The arrows are indicating the magnitude and the direction of the fluid velocity
  $U_f$.}
  \label{fig:avalanching}
\end{figure}

The snapshots of the arrangement of the sediment particles are shown in Fig.~\ref{fig:avalanching}.
When the grains stop moving, the repose angle is measured by using the steepest angle of the
profile~\citep{zhao14thesis}. The repose angle obtained in the present simulation is $19.8^{\circ}$.
This angle is consistent with the results obtained by using mono-dispersed spheres ($20.0^{\circ}$
in~\cite{goniva12irf} , $22.6^{\circ}$ in~\cite{zhao13coupled}). Note that the repose angles of
poly-dispersed or irregular sand particles are larger than that of mono-dispersed
spheres~\citep{zhao13coupled,zhao14thesis}. 

\begin{table}[htbp]
 \caption{Parameters used in the grain avalanching test.}
 \begin{center}
 \begin{tabular}{lcccc}
   \hline
   Domain dimensions                            &\\
   \qquad length, height, thickness ($L_x/d_p$, $L_y/d_p$, $L_z/d_p$) & $160\times40\times8$ \\
   \qquad length, height, thickness in mm ($L_x$, $L_y$, $L_z$) & $80\times20\times4$\\
   Mesh resolutions                             &\\                        
   \qquad length, height, thickness ($N_x$, $N_y$, $N_z$)  & $80\times20\times1$ \\
  Particle properties & \\
   \qquad total number                          & \multicolumn{4}{ c }{6,000}\\
   \qquad diameter $d_p$~[mm]                   & \multicolumn{4}{ c }{0.5}\\
   \qquad density $\rho_s$~[$\times 10^3~\mathrm{kg/m^3}$]  & \multicolumn{4}{ c }{2.5} \\
   \hline
  \end{tabular}
 \end{center}
 \label{tab:param-avalanching}
\end{table}
  
\end{document}